\documentclass[onecolumn,draftcls,12pt]{IEEEtran}
\usepackage[OT1]{fontenc}
\usepackage{amsmath,amsfonts}
\usepackage{algorithmic}
\usepackage{algorithm}
\usepackage{array}
\usepackage[caption=false,font=normalsize,labelfont=sf,textfont=sf]{subfig}
\usepackage{textcomp}
\usepackage{stfloats}
\usepackage{verbatim}
\usepackage{graphicx}
\usepackage{cite}
\usepackage{color}

\usepackage{mathrsfs}
\usepackage{amssymb}
\usepackage{bm}
\usepackage{epsfig}
\usepackage{theorem}
\usepackage{url}
\begin{document}

\title{A Unified Joint Optimization of Training Sequences and Transceivers Based on Matrix-Monotonic Optimization}

\author{Chengwen Xing, \textsl{Member}, \textsl{IEEE}, Tao Yu, Jinpeng Song, \textsl{Member}, \textsl{IEEE}, Zhong Zheng, \textsl{Member}, \textsl{IEEE}, Lian Zhao, \textsl{Senior Member}, \textsl{IEEE}, and Lajos Hanzo, \textsl{Life Fellow}, \textsl{IEEE}
\thanks{
	Manuscript received 7 Sept. 2022; revised 8 May 2023; accepted 19 May 2023. Date of current version 21  May 2023. 
	This work was supported in part by the National Natural Science Foundation of China under Grant 62231004 and 62071398, and the BIT  Research and  Innovation Promoting Project (Grant No. 2022YCXY045).
	L. Hanzo would like to acknowledge the financial support of the Engineering and Physical Sciences Research Council projects EP/W016605/1 and EP/X01228X/1 as well as of the European Research Council's Advanced Fellow Grant QuantCom (Grant No. 789028).
	The associate editor coordinating the review of this manuscript and approving it for publication was Dr. Xuanyu Cao. 
	\textit{(Corresponding author: Jinpeng Song.)}
}
\thanks{
	C. Xing, T. Yu, and Z. Zheng are with School of Information and Electronics, Beijing Institute of Technology, Beijing 100081, China (e-mail: xingchengwen@gmail.com, taoyu.bit@gmail.com, zhong.zheng@bit.edu.cn).
	
	J. Song is with School of Cyberspace Science and Technology, Beijing Institute of technology, Beijing 100081, China (e-mail: jinpeng@bit.edu.cn).
	
	L. Zhao is with the Department of Electrical, Computer and Biomedical Engineering, Ryerson University, Toronto, ON M5B 2K3, Canada (e-mail: l5zhao@ryerson.ca).
	
	L. Hanzo is with the Department of Electronics and Computer Science, University of Southampton, Southampton SO17 1BJ, UK (e-mail: lh@ecs.soton.ac.uk).
}
}


\maketitle

\begin{abstract}
Channel estimation and data transmission constitute the most fundamental functional modules of multiple-input multiple-output (MIMO) communication systems. The underlying key tasks corresponding to these modules are training sequence optimization and transceiver optimization. Hence, we jointly optimize the linear transmit precoder and the training sequence of MIMO systems using the metrics of their effective mutual information (MI), effective mean squared error (MSE), effective weighted MI, effective weighted MSE, as well as their effective generic Schur-convex and Schur-concave functions. Both statistical channel state information (CSI) and estimated CSI are considered at the transmitter in the joint optimization. A unified framework termed as joint matrix-monotonic optimization is proposed. Based on this, the optimal precoder matrix and training matrix structures can be derived for both CSI scenarios. Then, based on the optimal matrix structures, our linear transceivers and their training sequences can be jointly optimized. Compared to state-of-the-art benchmark algorithms, the proposed algorithms visualize the bold explicit relationships between the attainable system performance of our linear transceivers conceived and their training sequences, leading to implementation ready recipes. Finally, several numerical results are provided, which corroborate our theoretical results and demonstrate the compelling benefits of our proposed pilot-aided MIMO solutions.
\end{abstract}


\section{Introduction}
In the evolution of commercial cellular networks from 4G to next
generation solutions, multiple-input multiple-output (MIMO) technology
constitutes a salient milestone
\cite{Yang2015,Zheng2015,Xie2017,Wang2016,Wang2020,Ma2021,Wang2019,Wan2021,Fang2017},
leading to significant spectral and power efficiency
improvements. However, they require accurate channel state information
(CSI) \cite{Palomar03,XingTSP201501,Xie2016,Wu2019,Jiang2010,Xu2022}. Therefore, pilot-aided
channel estimation plays an important role in multi-antenna system
design
\cite{Hassibi2003,Yuan2018,You2015,Jin2016,Qi2015,Wu2020}.  Based on the
noise-contaminated pilot observations, the MIMO channel matrix can be
estimated by relying on performance metrics such as the least square error
\cite{Biguesh2006}, minimum mean squared error (MSE)
\cite{Soysal2010,Soysal2010-2}, mutual information
(MI)\cite{Coldrey2007,Coldrey2008}, etc. Then, by leveraging the
estimated CSI, MIMO transceivers can be designed for optimizing the
overall system performance
\cite{Pastore2016,Soysal2010,Soysal2010-2}.

The specific choice of the training sequence has a pivotal impact on
the performance of channel estimation for MIMO systems
\cite{Ma2017,Yuan2018},
and the key task of channel estimation is to recover the MIMO channel
matrix instead of a scalar parameter.
Hence, MIMO channel estimation constitutes a multiple-parameter
estimation problem, where the training sequence optimization relies
either on signal processing oriented
\cite{Wong2004,Shariati2014} or on information theoretic
metrics \cite{Gu2019}.  It has been shown based on
matrix-monotonic optimization \cite{XingTSP201501} that in fact a
diverse variety of training sequence optimization techniques relying
on heterogeneous performance metrics intrinsically aim for maximizing
a matrix-valued signal-to-noise ratio (SNR) during the channel estimation
procedure.

It is widely exploited that MIMO transceivers are capable of striking
a tradeoff between the spatial diversity gain and the spatial
multiplexing gain attained by the MIMO channels. When multiple data
streams are transmitted simultaneously from the source to destination,
the transceiver optimization constitutes a multi-objective
optimization problem \cite{Xing2021,Xing2021-2,Fei2017}.  Given the
CSI, diverse performance metrics, such as the capacity, or the sum MSE
of the estimation may be optimized \cite{Palomar03}. In order to
accommodate as many optimization objective functions (OFs) as
possible, majorization theory is used in \cite{Marshall79,Palomar03}
to formulate unified OFs~\footnote{Explicitly, we will
use the adjective ``unified'' upon referring to unifying diverse
OFs, and the adjective ``joint'' upon referring to
the optimization of both the training sequence and
precoder.}. Recently, the framework of matrix-monotonic optimization
was proposed for MIMO transceiver optimization relying on diverse OFs
\cite{XingTSP201501,Xing2021,Xing2021-2}. Again, as pointed out in
\cite{XingTSP201501}, the MIMO precoder optimization task is
reminiscent of maximizing a matrix-valued SNR in the data transmission
procedure.

When channel estimation is more accurate, at first right the illusion
might appear that the successive data transmission will have improved
performance. However, when either the pilot overload or the pilot
power is increased, either the effective throughput or the power - or
potentially both - have to be reduced for data transmission for the
sake of fair comparison. This tradeoff reveals that the joint
optimization of the transmit precoder (TPC) and of the training
sequence is of critical importance for MIMO systems. Hence, there is a
rich body of literature on joint training and transceiver optimization
\cite{Pastore2016,Soysal2010}, but in most of the related literature,
the joint optimization tends to rely on a single performance
metric. For example, joint TPC and training sequence optimization
relying on MI maximization was extensively studied in
\cite{Pastore2016,Soysal2010,Soysal2010-2}. However, using a joint
training sequence and transceiver weight optimization framework
relying on multiple performance metrics - rather than a single one,
like the MI in~\cite{Pastore2016,Soysal2010,Soysal2010-2} - has
important theoretical and practical benefits. Having said that, there
is a distinct paucity of literature on this topic - probably because
their joint optimization is much more challenging than that of its
separate counterparts. This becomes particularly challenging, when
both the time domain as well as the inter-element correlation of the
channel's transmit side are taken into account \cite{Pastore2016}. The
main challenge arises from the complex performance analysis of the
associated random matrix variables
\cite{Soysal2010,Soysal2010-2,Hassibi2003}. 
The distinct contributions of this paper are shown in Table~\ref{comparison}.

\begin{table}[h!]
  \centering
  \vspace{-3mm}
  \caption{ {Key contributions of this paper contrasted to the existing joint design of training sequences and MIMO transceivers with the power constrained}}
  \vspace{-2mm}
  \label{comparison}
  \begin{tabular}{| m{1.7cm} | m{0.9cm} | m{0.9cm} | m{3cm} |}
   \hline
    & [23]  & [25] & Our \\
   \hline\hline
   Transmit-side correlation channel model   & $\surd$ & $\surd$ & $\surd$ \\
   \hline
   Statistical CSI known at transmitter   & $\surd$ & $\surd$ & $\surd$ \\
   \hline
   Estimated CSI known at transmitter   &  &  & $\surd$ \\
   \hline
   OFs   & effective MI & different OFs & applicable to OFs in [25] and Schur-concave/convex
   functions \\
   \hline
   Maximum number of antennas in simulations   & 3 & 2 & 16 \\
   \hline
  \end{tabular}
  \vspace{-1mm}
\end{table}

In general MIMO communication systems, multi-dimensional signals have to be processed both in channel estimation and data transmission, where diverse tradeoffs must be struck. The matrix-version SNR is capable of accurately characterizing these tradeoffs.
Matrix-monotonic optimization handles matrix variables, where a positive semi-definite matrix is used as the performance metric. This optimization procedure exploits the monotonicity in the positive semi-definite matrix cone for deriving the optimal structures of the Pareto optimal matrix variables  \cite{XingTSP201501}. 
Generally speaking, as for MIMO systems based on matrix-monotonic optimization,  both the precoder optimization and training optimization aim for maximizing a matrix-version SNR instead of a scalar SNR. This general multi-objective optimization procedure offers substantial flexibility for mathematical derivations.
Based on the optimal structures derived, the optimization problems of MIMO-aided communications can be significantly simplified, resulting into traditional vector-valued optimization problems.

Moreover, a close scrutiny reveals that the nature of training optimization and of transceiver optimization is quite similar to each other \cite{XingTSP201501}.  Hence, we are inspired to investigate the joint optimization of the training sequence design and the TPC design based on the matrix-monotonic optimization framework of \cite{XingTSP201501}. A novel matrix-monotonic optimization technique termed as joint matrix-monotonic optimization is proposed for the joint optimization of the training sequence and of a linear TPC, which involves a pair of matrix variables, namely the training sequence matrix and linear TPC matrix.  Our main contributions are as follows:
\begin{itemize}

\item A whole suite of performance metrics is considered in the joint
  optimization of linear transceivers and training sequences,
  including the effective MI, effective sum MSE, effective weighted
  MI, effective weighted MSE, as well as the effective general
  Schur-convex and Schur-concave functions. Furthermore, a range of
  practical settings is considered. Specifically, for channel
  estimation, the power of the amplifier is limited, hence the
  estimation accuracy could be improved by increasing the length of
  the training sequence instead of increasing the power of the
  amplifiers.

\item For the joint optimization of linear TPCs and training
  sequences, the spatial correlation of antennas experienced at the
  transmitter is taken into account. Moreover, both statistical CSI
  and estimated CSI are discussed. More explicitly for the former
  scenario, the transmitter only has statistical CSI information,
  while the receiver relies on the estimated CSI. By contrast, for the
  latter case, the transmitter and the receiver have the same
  estimated CSI, i.e. the feedback channel between the transmitter and
  receiver is assumed to be perfect.

  {
\item Our matrix-monotonic optimization framework of
  \cite{XingTSP201501} derived for separate transceiver optimization
  or training optimization is extended here to the new joint
  matrix-monotonic optimization concept, which jointly optimizes our
  linear TPC as well as the training sequence, and judiciously allocates the system resources to optimize the effective performances of integrated pilot-aided MIMO communication systems. Based on our new joint
  matrix-monotonic optimization framework, the optimal structures of the linear
  TPC matrix and of the training sequence matrix are derived. Therefore,
  the resultant joint optimization problems can be significantly
  simplified. Although the original joint optimization problems are
  complex, the proposed solution is of appealingly low complexity,
  which provides valuable guidelines for practical system designs.
  }
\end{itemize}

\begin{figure}[t]
  \centering
  \includegraphics[width=7.2cm]{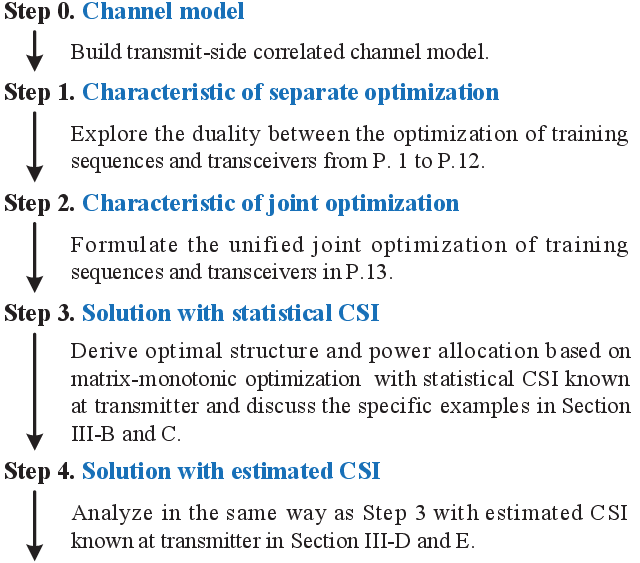}
  \vspace{-3mm}
  \caption{Flow of the mathematical analysis. }
  \vspace{-5mm}
  \label{flow}
\end{figure}

The rest of the paper is organized as follows. In Section~\ref{Section_Signal_Model}, the signal models of channel estimation and information transmission are provided, respectively. The general joint training sequence and linear transceiver optimization problem is formulated and solved in Section~\ref{Section_Joint_Optimization}. Specifically, the joint optimization of the training sequence and transceiver is investigated in Section~\ref{Section_Transmitter_Statistical} when the transmitter has only statistical CSI. By contrast, the transmitter relying on estimated CSI is investigated in Section~\ref{Section_Transmitter_Estimated_CSI}. Our simulation results are discussed in Section~\ref{Section_Simulation}, while our conclusions are offered in Section~\ref{Section_Conclusions}. Fig.~\ref{flow} illustrates the flow of mathematical analysis.

\noindent \textbf{Notation:} In order to clarify the following mathematical derivations, we introduce the notations, symbols and definitions used throughout this paper. Referring to the fundamental matrix operations, the symbols ${\boldsymbol{Z}}^{\rm{H}}$ and ${\boldsymbol{Z}}^{\rm{T}}$  denote the Hermitian transpose and the transpose of a general matrix ${\boldsymbol{Z}}$, respectively. The trace and determinant of a square matrix ${\boldsymbol{Z}}$ are denoted as  ${\rm{Tr}}({\boldsymbol{Z}})$ and $|{\boldsymbol{Z}}|$, respectively. For a positive semidefinite matrix $\bm{Z}$, the matrix ${\boldsymbol{Z}}^{\frac{1}{2}}$ or ${\boldsymbol{Z}}^{{1}/{2}}$  is the Hermitian square root of ${\boldsymbol{Z}}$, which is also positive semidefinite. The mathematical notation ${z^ + }$ represents $\max \{ 0,z\} $. For matrix decompositions in this paper, the notations ${\boldsymbol \Lambda} \searrow $ and  ${\boldsymbol \Lambda} \nearrow $ represent rectangular or square diagonal matrices with their diagonal elements sorted in decreasing and increasing order, respectively.

\section{Signal Models  for Channel Estimation and Data Transmission}
\label{Section_Signal_Model}

\begin{figure*}[!t]
  \hrulefill
  \normalsize
  \centering
  \includegraphics[width=18cm]{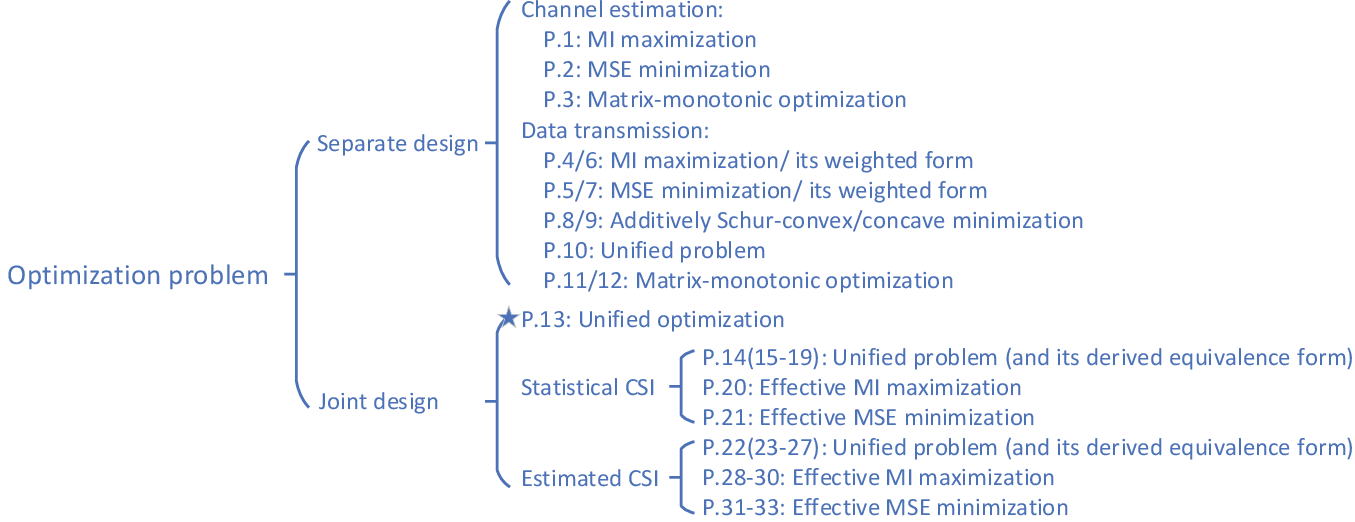}
  \vspace{-3mm}
  \caption{ {Overview of the OFs in the paper. }}
  \label{problem}
  \end{figure*}

A typical point-to-point MIMO system is considered. Specifically, an $N_{\rm{R}} \times N_{\rm{T}}$ MIMO channel $\bm{H}$ associated with transmit-side spatial antenna correlation is investigated, where the channel $\bm{H}$ can be formulated as in \cite{Pastore2016,Xing2021,Soysal2010}:
\begin{align}\label{Channel_Model}
{\bm{H}}={\bm{H}}_{\rm{W}}{\bm \Psi}^{1/2},
\end{align}where the entries of ${\bm{H}}_{\rm{W}}$ are independent and identically distributed (i.i.d.) Gaussian random variables having zero mean and unit variance. In the signal model (\ref{Channel_Model}), the positive definite matrix ${\bm \Psi}$ represents the transmit correlation matrix. Both channel estimation and data transmission are considered assuming the channel model (\ref{Channel_Model}), which are performed consecutively. In the first phase, training sequences are transmitted. Based on the noise-contaminated training signal observations, the channel matrix is estimated at the receiver. In the second phase, based on the estimated channel matrix, the transmit precoded signals are constructed and transmitted, which are then recovered at the receiver. 
{ 
In the following, we discuss the signal models of these two phases respectively. The connections and equivalences among different optimization metrics and OFs are illustrated in Fig.~\ref{problem}.
Firstly, we discuss different OFs and their unified matrix-monotonic optimizations of separate designs in channel estimation and data transmission respectively, in order to explore the common traits within them \cite{XingTSP201501}. More details on the differences and physical meanings among these OFs may be found in Appendix A. Then, we take the resource allocation into consideration and formulate a joint optimization problem relying on a unified framework in Section~\ref{Section_Joint_Optimization}, which is applicable to various OFs. Specifically, two scenarios are taken into consideration, which rely on either statistical or estimated CSI at the transmitter. The unified problem is derived, and the optimal structure is obtained. Then, some specific OFs are discussed and the corresponding solutions are provided.
}

\subsection{Channel Estimation and Training Optimization}

For the training based channel estimation, the training sequence $\bm{X}$ is transmitted to the receivers, yielding the received signal sequence of
\begin{align}
\label{Channel_Estimation_Model}
{\bm{Y}}={\bm{H}}{\bm{X}}+{\bm{N}},
\end{align} where $\bm{N}$ is the additive noise matrix at the destination. Based on the signal model in (\ref{Channel_Estimation_Model}), the key task is to recover the channel matrix $\bm{H}$ from the noise-contaminated observation $\bm{Y}$ as accurately as possible. For a linear channel estimator, the estimated channel equals \cite{Pastore2016}
\begin{align}
\label{Estimated_Channel}
{\bm{\widehat H}}={\bm{Y}}{\bm{G}}_{\rm{E}},
\end{align} where the estimated channel matrix ${\bm{\widehat H}}$ and the true channel matrix ${\bm{H}}$ satisfy
\begin{align}\label{Error_Model}
{\bm{H}}=&{\bm{\widehat H}}+ \Delta{\bm{H}},
\end{align} where $\Delta\bm{H}$ is the estimation error.
Then the MSE matrix of the corresponding channel estimation can be expressed as
\begin{align} \label{MMSE_Matrix}
\bm{E}_{\rm{MSE}}&= {\mathbb{E}}\{\Delta\bm{H}^{\rm{H}}\Delta\bm{H} \} \nonumber \\
&=\ ({\bm{I}}
-{\bm{X}}{\bm{G}}_{\rm{E}})^{\rm{H}}{\bm{R}}_{\rm{H}}
({\bm{I}}-{\bm{X}}{\bm{G}}_{\rm{E}})
+{\bm{G}}_{\rm{E}}^{\rm{H}}{\bm{R}}_{\rm{N}}{\bm{G}}_{\rm{E}},
\end{align}where the channel's correlation matrix ${\bm{R}}_{\rm{H}}$ and the noise covariance matrix ${\bm{R}}_{\rm{N}}$ are defined as
\begin{align}
{\bm{R}}_{\rm{H}}&={\mathbb{E}}\{{\bm{H}}^{\rm{H}}{\bm{H}}\}=N_{\rm{R}}{\bm \Psi}, \
\bm{R}_{\rm{N}} =\mathbb{E} \{\bm{N}^{\rm{H}}  \bm{N}\},
\end{align} where the scalar $N_{\rm{R}}$ denotes the number of receive antennas. In order to minimize $\bm{E}_{\rm{MSE}}$, the optimal ${\bm{G}}_{\rm{E}}$ can be chosen as
the linear minimum mean squared error (LMMSE) channel estimator as in  \cite{Kay93,Ding09}:
\begin{align}\label{LMMSE_Estimator}
{\bm{G}}_{{\rm{E}},{\text{Opt}}}=&({\bm{X}}^{\rm{H}}{\bm{R}}_{\rm{H}}{\bm{X}}+{\bm{R}}_{\rm{N}})^{-1}
{\bm{X}}^{\rm{H}}{\bm{R}}_{\rm{H}}.
\end{align}Therefore, the channel estimation MSE matrix $\bm{E}_{\rm{MSE}}$ in (\ref{MMSE_Matrix}) can be reformulated as
\begin{align}
\label{MSE_Matrix_Channel_Estimation}
\bm{E}_{\rm{MSE}}=&\mathbb{E}\{  \Delta{\bm{H}}^{\rm{H}} \Delta{\bm{H}} \} \nonumber \\
=&({\bm{R}}_{\rm{H}}^{-1}+{\bm{X}}{\bm{R}}_{\rm{N}}^{-1}{\bm{X}}^{\rm{H}})^{-1} \nonumber \\
=&N_{\rm{R}}\left({\bm{\Psi}}^{-1}+N_{\rm{R}}{\bm{X}}{\bm{R}}_{\rm{N}}^{-1}{\bm{X}}^{\rm{H}}\right)^{-1}
.
\end{align} Meanwhile, based on the LMMSE estimator in (\ref{LMMSE_Estimator}), the resultant channel estimation error $\Delta\bm{H}$ can be written in the following form \cite{Kay93}
\begin{align}
\Delta\bm{H}=\Delta\bm{H}_{\rm{W}}{\bm \Phi}^{1/2},
\end{align} where $\Delta\bm{H}_{\rm{W}}$ is a random matrix whose elements are i.i.d. Gaussian distributed with zero mean and unit variance, while
${\bm \Phi}$ is given in (\ref{MSE_Matrix_Channel_Estimation}) as
\begin{align}
{\bm \Phi}
=& \left({\bm{\Psi}}^{-1}+N_{\rm{R}}{\bm{X}}{\bm{R}}_{\rm{N}}^{-1}
{\bm{X}}^{\rm{H}}\right)^{-1}.
\end{align}
Moreover, upon applying the LMMSE channel estimator of (\ref{LMMSE_Estimator}), the correlation matrix of the estimated channel ${\mathbb{E}}\{\bm{\widehat H}^{\rm{H}}\bm{\widehat H}\}$ equals
\begin{align}
{\mathbb{E}}\{\bm{\widehat H}^{\rm{H}}\bm{\widehat H}\}
&=\! {\bm{R}}_{\rm{H}}\!-\! ({\bm{R}}_{\rm{H}}^{-1}\!+ \! {\bm{X}}{\bm{R}}_{\rm{N}}^{-1}{\bm{X}}^{\rm{H}})^{-1} \nonumber \\
&=\! N_{\rm{R}}{\bm{\Psi}}\! - \! N_{\rm{R}}\left({\bm{\Psi}}^{-1}\!+\!N_{\rm{R}}{\bm{X}}{\bm{R}}_{\rm{N}}^{-1}
{\bm{X}}^{\rm{H}}\right)^{-1}
\triangleq {\bm \Pi}.
\end{align}

 When the LMMSE channel estimator is adopted, the channel estimation MSE matrix $\bm{E}_{\rm{MSE}}$ is a function of the training sequence $\bm{X}$. Therefore, the performance of channel estimation may indeed be improved by optimizing the choice of the training sequence $\bm{X}$ \cite{XingTSP201501}. The most widely used performance metrics are the MI \cite{Gu2019} and MSE \cite{XingTSP201501}. In the following, these two classic training optimization techniques are reviewed in order to reveal the general optimal structure of the training sequence $\bm{X}$. As for MI maximization,
the corresponding training optimization problem is formulated as in  \cite{Coldrey2007}, \cite{Coldrey2008}, \cite{Gu2019}:
\begin{align}
\textbf{P. 1:} \ \ \max_{{\bm{X}}} \ \ & {\rm{log}}\ {\rm{det}}\big( {\bm{R}}_{\rm{H}}^{-1}+{\bm{X}}{\bm{R}}_{\rm{N}}^{-1}{\bm{X}}^{\rm{H}}\big), \ \nonumber \\
{\rm{s.t.}} \ & {\rm{Tr}}({\bm{X}}{\bm{X}}^{\rm{H}})\le P_{\rm{T}}T_{\rm{T}},
\end{align} where $P_{\rm{T}}$ and $T_{\rm{T}}$ are the power and the time interval of channel estimation, respectively, and $P_{\rm{T}}T_{\rm{T}}$ is the energy allocated to the channel estimation. 
{
This energy constraint can be viewed as the Frobenius norm of the training sequence matrix. 
It is worth noting that for matrix-valued training sequences there are multiple solutions that have the same Frobenius. As a result, the optimization aims for finding the optimal training sequence matrix in the whole feasible solution set.}
On the other hand, the training optimization aiming for minimizing the sum MSE can be formulated in the following form \cite{Wong2004}
\begin{align}
\textbf{P. 2:} \ \ \min_{{\bm{X}}} \ \ & {\rm{Tr}}[ ({\bm{R}}_{\rm{H}}^{-1}+{\bm{X}}{\bm{R}}_{\rm{N}}^{-1}
{\bm{X}}^{\rm{H}})^{-1}], \ \nonumber \\
{\rm{s.t.}} \ & {\rm{Tr}}({\bm{X}}{\bm{X}}^{\rm{H}})\le P_{\rm{T}}T_{\rm{T}}.
\end{align}Based on \textbf{P. 1} and \textbf{P. 2}, it may be inferred that using different performance metrics for MIMO channel estimation results in different optimal sequences, because multiple parameters have to be estimated in MIMO channel estimation. Different performance metrics result in different tradeoffs. Therefore, a general MIMO training optimization results in a multi-objective optimization problem.
From a multi-objective optimization perspective, all the optimal solutions of the training optimization associated with different performance metrics are the Pareto optimal\footnote{Pareto optimality refers to the fact that, in a multi-objective optimization problem, it is impossible to improve any of the objectives without degrading at least one of the other objectives.} solutions of the following matrix-monotonic optimization problem \cite{XingTSP201501}
\begin{align}
\textbf{P. 3:} \ \  \max_{\bm{X}} \ \ & {\bm{X}}{\bm{R}}_{\rm{N}}^{-1}{\bm{X}}^{\rm{H}}, \
 {\rm{s.t.}} \  {\rm{Tr}}({\bm{X}}{\bm{X}}^{\rm{H}}) \le P_{\rm{T}}T_{\rm{T}}.
\end{align} We would like to highlight that in the above optimization problem, the OF is a positive semi-definite matrix, which may be viewed as a matrix-valued SNR in the channel estimation procedure \cite{XingTSP201501}. Bearing in mind that the constraint in \textbf{P. 3} is unitary invariant, the matrix-monotonic optimization \textbf{P. 3} is of a vector optimization nature, which is a function of the vector consisting of the eigenvalues of ${\bm{X}}{\bm{R}}_{\rm{N}}^{-1}{\bm{X}}^{\rm{H}}$.
The different Pareto optimal solutions of \textbf{P. 3} achieve different levels of fairness among the different accuracies of the multiple estimated parameters. 
{
The channel estimation performance may be improved by optimizing the structure of the training sequence matrix depending on the availability of the channel's statistical information. From a practical implementation perspective, another design option is to gather more energy at the receiver instead of increasing the transmit power, because a practical power amplifier has a limited linear operating region. In other words, the channel estimation accuracy may also be improved by increasing the number of columns in $\bm{X}$, instead of increasing the amplitudes of the elements of $\bm{X}$. }
Based on the matrix-monotonic optimization framework of \cite{XingTSP201501}, we have the following conclusions for training optimization.

\noindent \textbf{Conclusion 1:} In MIMO channel estimation relying on an LMMSE estimator, the optimal training sequences satisfy the following structure
\begin{align}
{\bm{X}}_{\rm{opt}}={\bm{U}}_{\bm{X}}{\bm \Lambda}_{\bm{X}}{\bm{U}}_{{\bm{R}}_{\rm{N}}}^{\rm{H}},
\end{align} where the unitary matrix ${\bm{U}}_{{\bm{R}}_{\rm{N}}}$ is defined based on the eigenvalue decomposition (EVD) as
\begin{align}
{\bm{R}}_{\rm{N}}^{-1}={\bm{U}}_{{\bm{R}}_{\rm{N}}}{\bm \Lambda}_{{\bm{R}}_{\rm{N}}}^{-1}{\bm{U}}_{{\bm{R}}_{\rm{N}}}^{\rm{H}} \ \text{with} \ {\bm \Lambda}_{{\bm{R}}_{\rm{N}}}^{-1}\searrow.
\end{align}The unitary matrix ${\bm{U}}_{\bm{X}}$ is determined by the specific performance metrics. Finally, ${\boldsymbol \Lambda}_{\bm{X}}$ is a rectangular diagonal matrix, which is also determined by the specific performance metric.

\noindent \textbf{Conclusion 2:} Referring to ${\bm{U}}_{\bm{X}}$ for both MI maximization and MSE minimization, the optimal ${\bm{U}}_{\bm{X}}$ obeys \cite{XingTSP201501}
\begin{align}
{\bm{U}}_{\bm{X}}={\bm{U}}_{\bm{\Psi}},
\end{align}where the unitary matrix ${\bm{U}}_{\bm{\Psi}}$ is defined based on the following EVD:
\begin{align}
{\bm{\Psi}}={\bm{U}}_{\bm{\Psi}}{\bm \Lambda}_{\bm{\Psi}}{\bm{U}}_{\bm{\Psi}}^{\rm{H}} \  \text{with} \  {\bm \Lambda}_{\bm{\Psi}}\searrow.
\end{align}

Based on \textbf{Conclusion 2}, the structure ${\bm{U}}_{\bm{X}}={\bm{U}}_{\bm{\Psi}}$ is applied in the general joint optimization of the training and the transceiver in Section~\ref{Section_Joint_Optimization}. Moreover, it is also worth noting that at high SNRs\footnote{Channel estimation is performed invariably at high SNR.} the MI maximization \textbf{P. 1} has the optimal solution of ${\bm \Lambda}_{\bm{X}}=\bm{I}$, i.e., uniformly allocating the powers along spatial directions \cite{XingTSP201501}. By contrast, for MSE minimization, the optimal ${\bm \Lambda}_{\bm{X}}$ is the water-filling solution \cite{XingTSP201501}. Undoubtedly, uniform power allocation is the simplest scheme for training optimization, which can be adopted as a suboptimal power allocation scheme along the spatial directions.

\subsection{Transceiver Optimization}

Upon assuming that the estimated CSI is available at the transmitter with the aid of a perfect feedback channel, the transmitted signal is preprocessed by a channel-dependent TPC and the signal ${\bm{y}}$ received at the destination is of the following form \cite{Palomar03}:
\begin{align}\label{Signal_Model_Transceiver}
{\bm{y}}&={\bm{H}}{\bm{F}}{\bm{s}}+{\bm{n}},
\end{align}where ${\bm{F}}$ is the TPC matrix, ${\bm{s}}$ is the signal vector and ${\bm{n}}$ is the additive receiver noise having the covariance matrix ${\bm{R}}_{{\bm{n}}}=\sigma_{\rm{N}}^2{\bm{I}}$.
The covariance matrix of ${\bm{s}}$ is assumed to be an identity matrix, i.e., ${\mathbb{E}}\{{\bm{s}}{\bm{s}}^{\rm{H}}\}={\bm{I}}$. In data transmission, based on the channel estimation error model of (\ref{Error_Model}), the signal model of (\ref{Signal_Model_Transceiver}) used for data transmission is rewritten as
\begin{align}\label{Signal_Model_Data_T}
{\bm{y}}&={\bm{\widehat H}}{\bm{F}}{\bm{s}}+\Delta\bm{H}{\bm{F}}{\bm{s}}+{\bm{n}}
={\bm{\widehat H}}{\bm{F}}{\bm{s}}+\underbrace{\Delta\bm{H}_{\rm{W}}\bm{\Phi}^{\frac{1}{2}}
{\bm{F}}{\bm{s}}+{\bm{n}}}_{\triangleq \bm{v}},
\end{align}where $\bm{v}$ is the equivalent noise that consists of the additive noise and the channel estimation error. The covariance matrix $\bm{R}_{\bm{v}}$ of the equivalent noise $\bm{v}$ can be expressed as \cite{Pastore2016,Ding09}
\begin{align}
\bm{R}_{\bm{v}}=[\sigma_{\rm{N}}^2+{\rm{Tr}}(\bm{F}\bm{F}^{\rm{H}}\bm{\Phi})]\bm{I}.
\end{align} As for the transceiver optimization, the receiver has the information including the estimated channel matrix $\bm{\widehat H}$ and the covariance matrix $\bm{R}_{\bm{v}}$. Again, at the transmitter, we assume that either the estimated CSI ${\bm{\widehat H}}$ or only the statistical CSI given in (\ref{Channel_Model}) is available. In the following, both of these two cases are discussed.
Note that similarly to training optimization, for the MIMO TPC designs there are also diverse performance metrics that correspond to distinctly different OFs \cite{Palomar03,XingTSP201501}. 
In Appendix~\ref{Appendix}, the OFs considered in the following are derived in detail.

First, the TPC is designed by maximizing the capacity or MI. The corresponding optimization problem is given by \cite{Palomar03}
\begin{align}
\textbf{P. 4:}\!  \max_{{\bm{F}}}  & \mathbb{E}\!\left\{\!{\rm{log}}\det\!\left({\bm{I}}\!\!+\!\!{\bm{F}}^{\rm{ H}}{\bm{\widehat H}}^{\rm{H}}{\bm{R}}_{\bm{v}}^{-1}{\bm{\widehat H}}
{\bm{F}}\!\right)\!\right\} \! \nonumber \\
 {\rm{s.t.}} & \bm{R}_{\bm{v}}\!=\![\sigma_{\rm{N}}^2\!\!+\!\!{\rm{Tr}}(\bm{F}\bm{F}^{\rm{H}}\bm{\Phi})]\bm{I}, {\rm{Tr}}({\bm{F}}{\bm{F}}^{\rm{H}}) \!\le \! P_{\rm{D}},
\end{align}where $ P_{\rm{D}}$ is the maximum power during the data transmission. In \textbf{P. 4}, when the transmitter has the same CSI at the receiver, the expectation $\mathbb{E}\{\cdot\}$ can be removed. Otherwise, when only statistical CSI is available at the transmitter, $\mathbb{E}\{\cdot\}$ is performed over ${\bm{\widehat H}}$ for the optimization of ${\bm{F}}$. On the other hand, the TPC optimization aiming for minimizing the sum MSE may be written as \cite{Palomar03}
\begin{align}
\textbf{P. 5:} \! \min_{{\bm{F}}} & \mathbb{E}\!\left\{{\rm{Tr}}[\!({\bm{I}}\!+\!{\bm{F}}^{\rm{H}}{\bm{\widehat H}}^{\rm{H}}{\bm{R}}_{{\bm{v}}}^{-1}{\bm{\widehat H}}{\bm{F}}\!)^{-1}]\! \right\}\! \nonumber \\
 {\rm{s.t.}} & \bm{R}_{\bm{v}}\!=\![\sigma_{\rm{N}}^2\!+\!{\rm{Tr}}(\bm{F}\bm{F}^{\rm{H}}\bm{\Phi})]\bm{I}, {\rm{Tr}}({\bm{F}}{\bm{F}}^{\rm{H}}) \!\le\! P_{\rm{D}}.
\end{align} In order to extend the optimization \textbf{P. 4} to a general case, a weighted MI is formulated as \cite{Xing2021}
\begin{align}
\textbf{P. 6:} \! \max_{{\bm{F}}} & \mathbb{E}\!\left\{ \! {\rm{log}}\det\!\left(\!{\bm{I}}\!\!+\!\!{\bm{A}}^{\rm{H}}{\bm{F}}^{\rm{H}}{\bm{\widehat H}}^{\rm{H}}{\bm{R}}_{\bm{v}}^{-1}{\bm{\widehat H}}
{\bm{F}}{\bm{A}}\!\right)\! \right\} \! \nonumber \\
 {\rm{s.t.}} & \bm{R}_{\bm{v}}\!\!=\!\![\sigma_{\rm{N}}^2\!\!+\!\!{\rm{Tr}}(\bm{F}\bm{F}^{\rm{H}}\bm{\Phi})]\bm{I}, \!
 {\rm{Tr}}(\!{\bm{F}}{\bm{F}}^{\rm{H}}\!)\! \!\le\! \! P_{\rm{D}},
\end{align}where $\bm{A}$ is a complex matrix of appropriate dimensions. In \textbf{P. 6}, it can be observed that the weighting matrix $\bm{A}$ is applied to the matrix-valued SNR.
Similarly, the optimization problem of weighted MSE minimization can be written as \cite{XingTSP201501}
\begin{align}
\textbf{P. 7:} \!  \min_{{\bm{F}}} & \mathbb{E}\left\{ \! {\rm{Tr}}\![\!\bm{W}(\!{\bm{I}}\!+\!{\bm{F}}^{\rm{H}}{\bm{\widehat H}}^{\rm{H}}{\bm{R}}_{{\bm{v}}}^{-1}{\bm{\widehat H}}{\bm{F}})^{-1}\!]\! \right\} \nonumber \\
 {\rm{s.t.}} & \bm{R}_{\bm{v}}\!=\![\sigma_{\rm{N}}^2\!\!+\!\!{\rm{Tr}}(\bm{F}\bm{F}^{\rm{H}}\bm{\Phi})]\bm{I},\! {\rm{Tr}}({\bm{F}}{\bm{F}}^{\rm{H}})\! \le\! P_{\rm{D}},
\end{align} where the positive semidefinite matrix $\bm{W}$ is the weighting matrix.

Generally speaking, for linear transceiver optimization having additively Schur-convex objectives, the problem can be represented as \cite{Palomar03,XingTSP201501,Xing2021}
\begin{align}
\textbf{P. 8:}\! \min_{{\bm{F}}} & \mathbb{E}\!\left\{ \! f_{{\rm{Convex}}}^{\rm{A\!-\!Schur}}\!\left(\!{\bf{d}}\![\!({\bm{F}}^{\rm{H}}{\bm{\widehat H}}^{\rm{H}}
{\bm{R}}_{{\bm{v}}}^{-1}{\bm{\widehat H}}{\bm{F}}\!\!+\!\!{\bf{I}}\!)^{-1}\!]\! \right)\! \right\}\!\! \nonumber \\
 {\rm{s.t.}} & \bm{R}_{\bm{v}}\!\!=\!\![\sigma_{\rm{N}}^2\!\!+\!\!{\rm{Tr}}(\bm{F}\bm{F}^{\rm{H}}\bm{\Phi})]\bm{I}, \! {\rm{Tr}}({\bm{F}}{\bm{F}}^{\rm{H}})\!\!\le\!\! P_{\rm{D}},
\end{align} where $f_{{\rm{Convex}}}^{\rm{A\!-\!Schur}}(\cdot)$ is an additively Schur-convex function of the vector consisting of the diagonal elements of $({\bm{F}}^{\rm{H}}{\bm{\widehat H}}^{\rm{H}}
{\bm{R}}_{{\bm{v}}}^{-1}{\bm{\widehat H}}{\bm{F}}\!+\!{\bf{I}})^{-1}$. The symbol ${\bf{d}}(\bm{Z})$ represents the vector consisting of the diagonal elements of $\bm{Z}$.
On the other hand, the linear transceiver optimization having additively Schur-concave objective is given by \cite{Palomar03,XingTSP201501,Xing2021}
\begin{align}
\textbf{P. 9:}\! \min_{{\bm{F}}} &  \mathbb{E}\!\left\{ \! f_{{\rm{Concave}}}^{\rm{A\!-\!Schur}}\!\left(\!{\bf{d}}\![\!({\bm{F}}^{\rm{H}}{\bm{\widehat H}}^{\rm{H}}
{\bm{R}}_{{\bm{v}}}^{-1}{\bm{\widehat H}}{\bm{F}}\!\!+\!\!{\bm{I}}\!)^{-1}\!] \!\right)\! \right\}\! \nonumber \\
\!  {\rm{s.t.}} & \bm{R}_{\bm{v}}\!=\![\sigma_{\rm{N}}^2\!\!+\!\!{\rm{Tr}}\!(\bm{F}\bm{F}^{\rm{H}}\bm{\Phi})\!]\bm{I}, \!{\rm{Tr}}\!({\bm{F}}{\bm{F}}^{\rm{H}}\!)\!\le\! P_{\rm{D}},
\end{align} where $f_{{\rm{Concave}}}^{\rm{A\!-\!Schur}}(\cdot)$ is an additively Schur-concave function of the vector consisting of the diagonal elements of $({\bm{F}}^{\rm{H}}{\bm{\widehat H}}^{\rm{H}}
{\bm{R}}_{{\bm{v}}}^{-1}{\bm{\widehat H}}{\bm{F}}+{\bf{I}})^{-1}$.

In order to accommodate the OFs in \textbf{P. 4} to \textbf{P. 9}, the unified general optimization problem is formulated as follows
\begin{align}
\textbf{P. 10:} \max_{{\bm{F}}}  & f_{\rm{unified}}\!\left(\!{\bm{F}}^{\rm{H}}{\bm{\widehat H}}^{\rm{H}}\bm{R}_{\bm{v}}^{-1}{\bm{\widehat H}}{\bm{F}}\!\right)\! \nonumber \\
 {\rm{s.t.}} & \bm{R}_{\bm{v}}\!=\![\sigma_{\rm{N}}^2\!\!+\!\!{\rm{Tr}}\!(\bm{F}\bm{F}^{\rm{H}}\bm{\Phi})\!]\bm{I}, \!{\rm{Tr}}
 ({\bm{F}}{\bm{F}}^{\rm{H}})\!\le\! P_{\rm{D}}.
\end{align} It is worth noting that \textbf{P. 10} aims for maximizing a performance metric, while some optimization problems from the set of \textbf{P. 4} to \textbf{P. 9} aim for minimizing some performance metrics. In order to make the mathematical formulas consistent, when the optimization problem considered aims for minimizing a nonnegative function, its OF is replaced by its inverse function. When the optimization problem considered aims for minimizing a negative function, its OF is replaced by its negative counterpart.
It should be highlighted that the expectation operation is contained in \textbf{P. 10}, and has either different mathematical meanings when the transmitter has either estimated CSI or statistical CSI. Specifically, when the transmitter relies on estimated CSI, the TPC $\bm{F}$ is optimized based on a specific realization of ${\bm{\widehat H}}$ and then the expectation operation is applied over the resultant OF. On the other hand, when the transmitter has only statistical information, the TPC $\bm{F}$ is optimized over the whole distribution of ${\bm{\widehat H}}$, instead of a realization of ${\bm{\widehat H}}$.

As for the TPC optimizations from the set \textbf{P. 4} to \textbf{P. 9}, when the estimated CSI is available at the transmitter, the optimal solutions are the Pareto optimal solutions of the following matrix-monotonic optimization  \cite{XingTSP201501}
\begin{align}
\textbf{P. 11:}  \max_{{\bm{F}}} & {\bm{F}}^{\rm{H}}{\bm{\widehat H}}^{\rm{H}}{\bm{R}}_{\bm{v}}^{-1}{\bm{\widehat H}}{\bm{F}} \nonumber \\
 \ {\rm{s.t.}} & \bm{R}_{\bm{v}}\!=\![\sigma_{\rm{N}}^2 \!+\!{\rm{Tr}}(\bm{F}\bm{F}^{\rm{H}}\bm{\Phi})]\bm{I},   {\rm{Tr}}({\bm{F}}{\bm{F}}^{\rm{H}}) \!\le \! P_{\rm{D}}.
\end{align} On the other hand, when the transmitter only has statistical information, the optimal solutions of \textbf{P. 4} to \textbf{P. 9} aim for maximizing the distribution of ${\bm{F}}^{\rm{H}}{\bm{\widehat H}}^{\rm{H}}{\bm{R}}_{\bm{v}}^{-1}{\bm{\widehat H}}{\bm{F}}$.
When ${\bm{\widehat H}}$ has only column correlations, maximizing the distribution of a random positive semidefinite matrix is equivalent to maximizing its expectation, i.e., ${\bm{F}}^{\rm{H}}\mathbb{E}\{{\bm{\widehat H}}^{\rm{H}}{\bm{R}}_{\bm{v}}^{-1}{\bm{\widehat H}}\}{\bm{F}}$ \cite{TrainingXing}. This is because when the distribution is maximized, for any given realization in the original distribution, it is always possible to find a realization from the optimized distribution, which is larger than the original realization and has the same probability density function (pdf). The detailed proof can be found in \cite{TrainingXing}.
As a result, when ${\bm{\widehat H}}$ has only column correlations and  the transmitter has only statistical CSI, the optimal solutions of \textbf{P. 4} to \textbf{P. 9} are the Pareto optimal solutions of the following matrix-monotonic optimization \cite{Xing2021}
\begin{align}
\textbf{P. 12:}  \max_{{\bm{F}}} & {\bm{F}}^{\rm{H}}\mathbb{E}\{{\bm{\widehat H}}^{\rm{H}}{\bm{R}}_{\bm{v}}^{-1}{\bm{\widehat H}}\}{\bm{F}} \ \nonumber \\
{\rm{s.t.}} & \bm{R}_{\bm{v}} \!=\![\sigma_{\rm{N}}^2\!+\!{\rm{Tr}}(\bm{F}\bm{F}^{\rm{H}}\bm{\Phi})]\bm{I}, {\rm{Tr}}({\bm{F}}{\bm{F}}^{\rm{H}})\!\le \! P_{\rm{D}}.
\end{align} \textbf{P. 12} can be viewed as the optimization of the distribution of ${\bm{F}}^{\rm{H}}{\bm{\widehat H}}^{\rm{H}}{\bm{R}}_{\bm{v}}^{-1}{\bm{\widehat H}}{\bm{F}}$. When multiple data streams are transmitted, the MIMO TPC optimization is a multi-objective optimization problem. Generally speaking, there is no solution that is optimal for all OFs.
It is plausible that the channel estimation procedure has a grave impact on the TPC optimization. Note that in \textbf{P. 10} and \textbf{P. 11}, the positive definite matrix $\bm{\Phi}$  is a function of the training sequence $\bm{X}$. Moreover, it is also worth highlighting that the communication resources are limited, i.e., $(T-T_{\rm{T}})P_{\rm{D}}+P_{\rm{T}}T_{\rm{T}}\le E_{\rm{total}}$, where $T$ is the channel's coherence time. The more resources are allocated to channel estimation, the more accurately the channel matrix can be estimated, but leaving less resources for information transmissions. As a result, there exist tradeoffs between the optimizations of \textbf{P. 3} and \textbf{P. 10}. The focus of our work is hence to jointly optimize these two matrix-monotonic optimization problems.

\section{Unified Optimization of the Training Sequence and the TPC}
\label{Section_Joint_Optimization}

Based on the discussions in Section~\ref{Section_Signal_Model}, when having perfect CSI at the transmitter, the matrix-monotonic optimization problem \textbf{P. 10} can be simplified to the following matrix-monotonic optimization \cite{XingTSP201501}
\begin{align}
\label{MM_Precoder}
\max_{\bm{F}} \ \ & {\bm{F}}^{\rm{H}}{\bm{H}}^{\rm{H}}{\bm{R}}_{\bm{n}}^{-1}{\bm{H}}{\bm{F}},
\ {\rm{s.t.}} \   {\rm{Tr}}({\bm{F}}{\bm{F}}^{\rm{H}}) \le P_{{\rm{D}}},
\end{align} which is of the same form as \textbf{P. 3}. 
{
They try to recover the received signals from the noise-contaminated observations of the known training sequence for channel estimation and for data transmission by relying on the known channel matrix, respectively. This fact reveals that there exists a conceptual duality between the TPC optimization and training sequence optimization \cite{XingTSP201501}, albeit
the constraints in \textbf{P. 3} and \textbf{P. 10} have slightly different physical meanings from a physical perspective. }
Specifically, the constraint in \textbf{P. 3} guarantees having limited total energy for channel estimation. On the other hand, the constraint in \textbf{P. 10} assures that the maximum power is lower than a specific threshold.

\subsection{Unified Optimization Problems}

For the joint optimization of the training sequence $\bm{X}$ and the TPC $\bm{F}$, a joint performance metric is a function of both $\bm{X}$ and $\bm{F}$. Again, here the training optimization is carried out by only using the statistical CSI. By contrast, the TPC optimization is carried out by using either the statistical CSI or the estimated CSI. As a result, for the joint optimization considered, the performance metric should be an average performance that is independent of the instantaneous CSI.
The corresponding unified joint optimization of the training sequence and the TPC is written in the following form
\begin{align}
 \textbf{P. 13:} \  \max_{{\bm{F}},{\bm{X}},T_{\rm{T}}} \ \ & \frac{T-T_{\rm{T}}}{T} f_{\rm{unified}}\left(\frac{{\bm{F}}^{\rm{H}}{\bm{\widehat H}}^{\rm{H}}{\bm{\widehat H}}{\bm{F}}}{\sigma_{\rm{N}}^2+{\rm{Tr}}({\bm \Phi}{\bm{F}}{\bm{F}}^{\rm{H}})}\right)  \nonumber \\
 \ {\rm{s.t.}} \  & \bm{\Phi}=\left({\bm{\Psi}}^{-1}+N_{\rm{R}}{\bm{X}}{\bm{R}}_{\bm{N}}^{-1}
{\bm{X}}^{\rm{H}}\right)^{-1} \nonumber \\
& {\rm{Tr}}(\bm{F}\bm{F}^{\rm{H}}) \le  P_{\rm{D}}, \ {\rm{Tr}}(\bm{X}\bm{X}^{\rm{H}}) \le P_{\rm{T}}T_{\rm{T}},  \nonumber \\
& (T-T_{\rm{T}})P_{\rm{D}}+P_{\rm{T}}T_{\rm{T}}\le E_{\rm{total}}.
\end{align} In contrast to \textbf{P. 10}, there are three optimization variables in \textbf{P. 13}, namely ${\bm{F}}$, ${\bm{X}}$, and $T_{\rm{T}}$. The discrete time interval $T_{\rm{T}}$ is the training length, which must satisfy $1\le T_{\rm{T}} \le T\!-\!1$. The scalar $T_{\rm{T}}$ determines the amount of resources allocated to channel estimation and hence also the resources left for data transmission. After introducing $T_{\rm{T}}$ in the OF, \textbf{P. 13} aims for striking a compelling tradeoff between the channel estimation accuracy and data transmission efficiency, which is distinctly different from \textbf{P. 10}. In order to boldly differentiate the traditional designs given by \textbf{P. 10}, the performance metrics in \textbf{P. 13} given by \textbf{P. 4} to \textbf{P. 9} are termed as the effective metrics.
For example, when considering the MI maximization problem \textbf{P. 4}, \textbf{P. 13} aims for maximizing the \textit{effective MI} (termed as achievable rate in \cite{Pastore2016}) instead of the original MI. The corresponding OF is
\begin{align}\label{Objective_Effective_MI}
  &\frac{T-T_{\rm{T}}}{T}f_{\rm{unified}}\left(\frac{{\bm{ F}}^{\rm{H}}\bm{\widehat H}^{\rm{H}}{\bm{\widehat H}} {\bm{F}}}{\sigma_{\rm{N}}^2+{\rm{Tr}}({\bm \Phi}{\bm{F}}{\bm{F}}^{\rm{H}})}\right) \nonumber \\
  =&\frac{T-T_{\rm{T}}}{T}
  {\mathbb{E}\left\{{\rm{log \ det}}\left(\bm{I}
  +\frac{{\bm{F}}^{\rm{H}}{\bm{\widehat H}}^{\rm{H}}{\bm{\widehat H}}{\bm{F}}}{\sigma_{\rm{N}}^2+{\rm{Tr}}({\bm \Phi}{\bm{F}}{\bm{F}}^{\rm{H}})}\right)\right\}}.
\end{align} The OF in (\ref{Objective_Effective_MI}) simultaneously
maximizes the MI and data transmission time interval $T-T_{\rm{T}}$.

As discussed above, for the optimization problems \textbf{P. 4} to \textbf{P. 9}, there are several ones aiming for minimizing a specific performance metric. When these kinds of optimizations are unified in \textbf{P. 13}, the OFs are replaced by their corresponding inverse functions. For example,
for the weighted MSE minimization of \textbf{P. 7},
the corresponding OF of \textbf{P. 13} is formulated as the reciprocal of the weighted MSE, i.e.,
\begin{align}
&\frac{T-T_{\rm{T}}}{T}f_{\rm{unified}}\left(\frac{{\bm{ F}}^{\rm{H}}\bm{\widehat H}^{\rm{H}}{\bm{\widehat H}} {\bm{F}}}{\sigma_{\rm{N}}^2+{\rm{Tr}}({\bm \Phi}{\bm{F}}{\bm{F}}^{\rm{H}})}\right) \nonumber \\
=&\frac{T-T_{\rm{T}}}{T}
\frac{1}{\mathbb{E}\left\{{\rm{Tr}}\left[\bm{W}\left(\bm{I}
+\frac{{\bm{F}}^{\rm{H}}{\bm{\widehat H}}^{\rm{H}}{\bm{\widehat H}}{\bm{F}}}{\sigma_{\rm{N}}^2+{\rm{Tr}}({\bm \Phi}{\bm{F}}{\bm{F}}^{\rm{H}})}\right)^{-1}\right]\right\}}.
\end{align} Here, similar to the concept of effective MI, the term
\begin{align}
\frac{T}{T-T_{\rm{T}}}\mathbb{E}\left\{{\rm{Tr}}\left[\bm{W}\left(\bm{I}
+\frac{{\bm{F}}^{\rm{H}}{\bm{\widehat H}}^{\rm{H}}{\bm{\widehat H}}{\bm{F}}}{\sigma_{\rm{N}}^2+{\rm{Tr}}({\bm \Phi}{\bm{F}}{\bm{F}}^{\rm{H}})}\right)^{-1}\right]\right\}
\end{align}may be viewed as the effective weighted MSE. Minimizing the effective MSE aims for simultaneously minimizing the traditional MSE and maximizing the data transmission time interval.

\subsection{Joint Optimization for Statistical CSI at the Transmitter}
\label{Section_Transmitter_Statistical}

In this section, we investigate the case in which the transmitter only has statistical CSI but the destination has more accurate estimated CSI. Here we would like to highlight that in contrast to most of the existing studies on transceiver optimization \cite{Xing2021}, in the joint optimization of a linear TPC and training sequence, only the average performances are considered instead of their counterparts relying on the instantaneous CSI. Strictly speaking, only the upper bounds of the system performance are investigated.
Following the rationale in \cite{Pastore2016}, the expectation operations are moved into the OFs of \textbf{P. 4} to \textbf{P. 9}. The resultant joint optimization problem becomes
\begin{align}
 \textbf{P. 14:} \ \  \max_{{\bm{F}},{\bm{X}},T_{\rm{T}}} \ \ & \frac{T-T_{\rm{T}}}{T}f_{\rm{unified}}\left(\frac{{\bm{ F}}^{\rm{H}}\mathbb{E}\{\bm{\widehat H}^{\rm{H}}{\bm{\widehat H}} \}{\bm{F}}}{\sigma_{\rm{N}}^2+{\rm{Tr}}({\bm \Phi}{\bm{F}}{\bm{F}}^{\rm{H}})}\right) \nonumber \\
 \ {\rm{s.t.}} \ \ & \bm{\Phi}=\left({\bm{\Psi}}^{-1}+N_{\rm{R}}{\bm{X}}{\bm{R}}_{\bm{N}}^{-1}
{\bm{X}}^{\rm{H}}\right)^{-1} \nonumber \\
& {\rm{Tr}}(\bm{F}\bm{F}^{\rm{H}}) \le  P_{\rm{D}}, \ {\rm{Tr}}(\bm{X}\bm{X}^{\rm{H}}) \le P_{\rm{T}}T_{\rm{T}}, \ \nonumber \\
&(T-T_{\rm{T}})P_{\rm{D}}+P_{\rm{T}}T_{\rm{T}}\le E_{\rm{total}}.
\end{align}
In the OF of \textbf{P. 14}, the average matrix-valued SNR becomes:
\begin{align}
 & \frac{{\bm{ F}}^{\rm{H}}\mathbb{E}\{\bm{\widehat H}^{\rm{H}}{\bm{\widehat H}} \}{\bm{F}}}{\sigma_{\rm{N}}^2+{\rm{Tr}}({\bm \Phi}{\bm{F}}{\bm{F}}^{\rm{H}})}= \frac{{\bm{ F}}^{\rm{H}}\bm{\Pi}{\bm{F}}}{\sigma_{\rm{N}}^2+{\rm{Tr}}({\bm \Phi}{\bm{F}}{\bm{F}}^{\rm{H}})}, \end{align}
with $ \bm{\Pi}=N_{\rm{R}}{\bm{\Psi}}\!-\!N_{\rm{R}}{\bm \Phi}$.

It can be concluded that \textbf{P. 14} maximizes this matrix-valued SNR. As a result, our joint matrix-monotonic optimization is formulated as
\begin{align}\label{Bi-Matrix-Monotonic}
 \textbf{P. 15:} \ \max_{{\bm{F}},{\bm{X}}} \  & \frac{{\bm{ F}}^{\rm{H}}\bm{\Pi}{\bm{F}}}{\sigma_{\rm{N}}^2+{\rm{Tr}}({\bm \Phi}{\bm{F}}{\bm{F}}^{\rm{H}})}\nonumber \\
 {\rm{s.t.}}  &    \bm{\Pi}\!=\!N_{\rm{R}}{\bm{\Psi}}\!-\!N_{\rm{R}}{\bm \Phi},
\bm{\Phi}\!=\!\!\left({\bm{\Psi}}^{-1}\!\!+\!N_{\rm{R}}{\bm{X}}{\bm{R}}_{\bm{N}}^{-1}
{\bm{X}}^{\rm{H}} \!\right)^{\!\!-1}\!\!, \nonumber \\
& {\rm{Tr}}(\bm{F}\bm{F}^{\rm{H}}) \le  P_{\rm{D}}, \ {\rm{Tr}}(\bm{X}\bm{X}^{\rm{H}}) \le P_{\rm{T}}T_{\rm{T}}, \ \nonumber \\
&(T-T_{\rm{T}})P_{\rm{D}}+P_{\rm{T}}T_{\rm{T}}\le E_{\rm{total}}.
\end{align}
Explicitly this joint matrix-monotonic optimization is intrinsically different from traditional single matrix-monotonic optimization applied separately for transceiver design and training design, since a pair of optimization variables, namely the linear TPC $\bm{F}$ and the training sequence $\bm{X}$, are jointly optimized. The optimal solutions of \textbf{P. 14} belong to the Pareto-optimal solution set of \textbf{P. 15}. We would like to point out that \textbf{P. 13} optimizes the distribution of ${\bm{F}}^{\rm{H}}{\bm{\widehat H}}^{\rm{H}}{\bm{R}}_{\bm{v}}^{-1}{\bm{\widehat H}}{\bm{F}}$ by relying on the statistical CSI available at transmitter.
Provided that ${\bm{\widehat H}}$ has only column correlations, there exists an equivalence between optimizing the distribution of a random positive semidefinite matrix and maximizing its expectation, i.e., ${\bm{F}}^{\rm{H}}\mathbb{E}\{{\bm{\widehat H}}^{\rm{H}}{\bm{R}}_{\bm{v}}^{-1}{\bm{\widehat H}}\}{\bm{F}}$ \cite{TrainingXing}. As a result,
in \textbf{P. 14}, there is an approximation, since the expectation operation is moved into the OF. However, based on our discussion concerning \textbf{P. 12}, this approximation does not change the fact that the optimal solutions of \textbf{P. 13} belong to the Pareto-optimal solution set of \textbf{P. 15}.  We can conjecture that the approximation in \textbf{P. 14} might change the correspondence between the optimal solutions of \textbf{P. 13} and the Pareto-optimal solutions of \textbf{P. 15}.

It is worth noting that the constraint ${\rm{Tr}}({\bm{F}}{\bm{F}}^{\rm{H}})\le P_{\rm{D}}$ is equivalent to the following constraint
\begin{align}
\label{power_constraint_1}
\sigma_{\rm{N}}^2{\rm{Tr}}({\bm{F}}{\bm{F}}^{\rm{H}}) \le \sigma_{\rm{N}}^2P_{\rm{D}},
\end{align} which is also equivalent to the following one
\begin{align}
\label{power_constraint_2}
\sigma_{\rm{N}}^2{\rm{Tr}}({\bm{F}}{\bm{F}}^{\rm{H}})\!+\!P_{\rm{D}}{\rm{Tr}}({\bm \Phi}{\bm{F}}{\bm{F}}^{\rm{H}})\!\le\! \sigma_{\rm{N}}^2P_{\rm{D}}\!+\!P_{\rm{D}}{\rm{Tr}}({\boldsymbol \Phi}{\bm{F}}{\bm{F}}^{\rm{H}}).
\end{align}The power constraint (\ref{power_constraint_2}) is equivalent furthermore to the following constraint
\begin{align}
\label{power_constraint_3}
\frac{{\rm{Tr}}[(\sigma_{\rm{N}}^2{\bm{I}}+P_{\rm{D}}{\boldsymbol \Phi}){\bm{F}}{\bm{F}}^{\rm{H}}]}{\sigma_{\rm{N}}^2+{\rm{Tr}}({\boldsymbol \Phi}{\bm{F}}{\bm{F}}^{\rm{H}})} \le P_{\rm{D}}.
\end{align} Here, it is worth noting that the equivalence implies that, (\ref{power_constraint_3}) can be proved and vice versa from (\ref{power_constraint_1}). Therefore, based on the equivalence between (\ref{power_constraint_1}) and (\ref{power_constraint_3}), the joint matrix-monotonic optimization problem \textbf{P. 15} is equivalent to the following one
\begin{align}
\label{Matrix_Monotonic_AA}
 \textbf{P. 16:}  \max_{{\bm{F}},{\bm{X}}}  &\frac{{\bm{ F}}^{\rm{H}}{\bm \Pi}{\bm{F}}}{\sigma_{\rm{N}}^2+{\rm{Tr}}({\bm \Phi}{\bm{F}}{\bm{F}}^{\rm{H}})}\nonumber \\
{\rm{s.t.}}  & \bm{\Pi}\!=\!\!N_{\rm{R}}{\bm{\Psi}}\!-\!N_{\rm{R}}{\bm \Phi},
\bm{\Phi}\!=\!\!\left({\bm{\Psi}}^{-1}\!+\!N_{\rm{R}}{\bm{X}}{\bm{R}}_{\rm{N}}^{-1}
{\bm{X}}^{\rm{H}}\right)^{-1}\!\!,   \nonumber \\
&{\rm{Tr}}(\bm{X}\bm{X}^{\rm{H}}) \!\le\! P_{\rm{T}}T_{\rm{T}},
\frac{{\rm{Tr}}[(\sigma_{\rm{N}}^2{\bm{I}}\!+\!P_{\rm{D}}{\boldsymbol \Phi}){\bm{F}}{\bm{F}}^{\rm{H}}]}{\sigma_{\rm{N}}^2+{\rm{Tr}}({\boldsymbol \Phi}{\bf{F}}{\bf{F}}^{\rm{H}})} \!\!\le\! P_{\rm{D}},  \nonumber \\
& (T-T_{\rm{T}})P_{\rm{D}}+P_{\rm{T}}T_{\rm{T}}\le E_{\rm{total}}.
\end{align} Upon defining the following new matrix variable
\begin{align}
\label{Definition_F}
{\bm{\widetilde F}}={(\sigma_{\rm{N}}^2{\bm{I}}+P_{\rm{D}}{\bm \Phi})^{1/2}}{\left[{\sigma_{\rm{N}}^2+{\rm{Tr}}({\bm \Phi}{\bm{F}}{\bm{F}}^{\rm{H}})}\right]^{-1/2}}\!{\bm{F}},
\end{align}the optimization problem \textbf{P. 16} can be simplified into the following one
\begin{align}
 \textbf{P. 17:} \ \max_{{\bm{\widetilde F}},\bm{X}} \ \ & {\bm{\widetilde F}}^{\rm{H}}(\sigma_{\rm{N}}^2{\bm{I}}+P_{\rm{D}}{\boldsymbol \Phi})^{-1/2}(N_{\rm{R}}\bm{\Psi}-N_{\rm{R}}{\bm{\Phi}}) \nonumber \\
 & \times(\sigma_{\rm{N}}^2{\bm{I}}+P_{\rm{D}}{\boldsymbol \Phi})^{-1/2}{\bm{\widetilde F}}\nonumber \\
 \ {\rm{s.t.}} \ &\bm{\Phi}=\left({\bm{\Psi}}^{-1}+N_{\rm{R}}{\bm{X}}{\bm{R}}_{\rm{N}}^{-1}
{\bm{X}}^{\rm{H}}\right)^{-1} \nonumber \\
& {\rm{Tr}}({\bm{\widetilde F}}{\bm{\widetilde F}}^{\rm{H}})\le P_{\rm{D}}, \ {\rm{Tr}}(\bm{X}\bm{X}^{\rm{H}}) \le P_{\rm{T}}T_{\rm{T}}, \nonumber \\
& (T-T_{\rm{T}})P_{\rm{D}}+P_{\rm{T}}T_{\rm{T}}\le E_{\rm{total}}.
\end{align}Based on the fundamental results of matrix-monotonic optimization \cite{XingTSP201501,Xing2021}, we have the following theorem concerning the optimal ${\bm{\widetilde F}}$.

\noindent \textbf{Theorem 1:} The Pareto-optimal solution of ${\bm{\widetilde F}}$ for \textbf{P. 17} satisfies the following structure
\begin{align}
{\bm{\widetilde F}}_{\rm{opt}}={\bm{V}}_{\bm{\mathcal{H}}}{\bm \Lambda}_{\bm{\widetilde F}}{\bm{U}}_{\bm{\widetilde F}}^{\rm{H}},
\end{align}where ${\bm{V}}_{\bm{\mathcal{H}}}$ and  ${\bm{U}}_{\bm{\widetilde F}}$ are unitary matrices and ${\bm \Lambda}_{\bm{\widetilde F}}$ is a diagonal matrix. The unitary matrix ${\bm{V}}_{\bm{\mathcal{H}}}$ is defined based on the following SVD
\begin{align}
(N\bm{\Psi}\!-\!N{\bm{\Phi}})^{\frac{1}{2}}(\sigma_{\rm{N}}^2{\bm{I}}\!+\!P_{\rm{D}}{\boldsymbol \Phi})^{-\frac{1}{2}}\!=\!{\bm{U}}_{\bm{\mathcal{H}}}{\bm \Lambda}_{\bm{\mathcal{H}}}
{\bm{V}}_{\bm{\mathcal{H}}}^{\rm{H}} \ \text{with} \ {\bm \Lambda}_{\bm{\mathcal{H}}} \searrow.
\end{align} On the other hand, the unitary matrix ${\bm{U}}_{\bm{\widetilde F}}$ is determined by the specific OFs \cite{XingTSP201501}. In the following, the optimal solutions of ${\bm{U}}_{\bm{\widetilde F}}$ are enumerated briefly and the detailed derivations are provided in \cite{XingTSP201501}.

\noindent \textbf{Conclusion 3:} For the optimization problems \textbf{P. 4} to \textbf{P. 9}, the corresponding optimal values of ${\bm{U}}_{\bm{\widetilde F}}$ are listed as follows:
\begin{align}
& \textbf{P. 4:} \  {\bm{U}}_{{\bm{\widetilde F}},{\rm{opt}}}\!=\!{\bm{U}}_{\rm{Arb}}; \ \textbf{P. 5:} \  {\bm{U}}_{{\bm{\widetilde F}},{\rm{opt}}}\!=\!{\bm{I}}; \ \textbf{P. 6:} \  {\bm{U}}_{{\bm{\widetilde F}},{\rm{opt}}}\!=\!{\bm{U}}_{\bm{A}}; \nonumber \\
&\textbf{P. 7:} \  {\bm{U}}_{{\bm{\widetilde F}},{\rm{opt}}}\!=\!
{\bm{U}}_{\bm{W}}; \ \textbf{P. 8:} \  {\bm{U}}_{{\bm{\widetilde F}},{\rm{opt}}}\!=\!{\bm{U}}_{\rm{DFT}}; \ \textbf{P. 9:} \  {\bm{U}}_{{\bm{\widetilde F}},{\rm{opt}}}\!\!=\!\!{\bm{I}},
\end{align}where the unitary matrices ${\bm{U}}_{\bm{A}}$ and ${\bm{U}}_{\bm{W}}$ are defined based on the following EVDs
\begin{align}
& \bm{A}\bm{A}^{\rm{H}}=\bm{U}_{\bm{A}}\bm{\Lambda}_{\bm{A}}\bm{U}_{\bm{A}}^{\rm{H}}  \ \text{with} \  {\bm \Lambda}_{\bm{A}} \searrow, \nonumber \\
& \bm{W}=\bm{U}_{\bm{W}}\bm{\Lambda}_{\bm{W}}\bm{U}_{\bm{W}}^{\rm{H}}  \ \text{with} \  {\bm \Lambda}_{\bm{W}} \searrow.
\end{align} Moreover, ${\bm{U}}_{\rm{Arb}}$ denotes an arbitrary unitary matrix of appropriate dimensions and ${\bm{U}}_{\rm{DFT}}$ represents a DFT unitary matrix of suitable dimensions, respectively.

According to the definition of ${\bm{\widetilde F}}$ in (\ref{Definition_F}), the following equality always holds
\begin{align}
(\sigma_{\rm{N}}^2{\bm{I}}+P_{\rm{D}}{\bm \Phi})^{-1/2}{\bm{\widetilde F}}=\left[{\sigma_{\rm{N}}^2+{\rm{Tr}}({\bm \Phi}{\bm{F}}{\bm{F}}^{\rm{H}})}\right]^{-1/2}{\bm{F}},
\end{align} based on which the following equality can be derived
\begin{align}
\frac{1}{{\sigma}_{\rm{N}}^2\!+\!{\rm{Tr}}\!(\!{\bm\Phi}{\bm{F}}{\bm{F}}^{\rm{H}})}
\!\!&=\!\!\frac{{\rm{Tr}}\![\!(\sigma_{\rm{N}}^2{\bm{I}}\!+\!P_{\rm{D}}{\bm \Phi})^{-\frac{1}{2}}{\bm{\widetilde F}}{\bm{\widetilde F}}^{\rm{H}}(\sigma_{\rm{N}}^2{\bm{I}}\!+\!P_{\rm{D}}{\bm\Phi}\!)^{-\frac{1}{2}}\!]}{{\rm{Tr}}({\bm{F}}{\bm{F}}^{\rm{H}})}\!\! \nonumber \\
&=\!\!\frac{{\rm{Tr}}[(\sigma_{\rm{N}}^2{\bm{I}}\!+\!P_{\rm{D}}{\bm \Phi})^{-1}{\bm{\widetilde F}}{\bm{\widetilde F}}^{\rm{H}}]}{P_{\rm{D}}},
\end{align}where the second equality is due to the fact that for the optimal ${\bm{F}}$, we have ${\rm{Tr}}({\bm{F}}{\bm{F}}^{\rm{H}})=P_{\rm{D}}$.
Finally, when the optimal solution ${\bm{\widetilde F}}$ has been computed, the optimal ${\bm{F}}$ equals
\begin{align}
{\bm{F}}_{\rm{opt}}\!=\!\sqrt{\frac{P_{\rm{D}}}{{\rm{Tr}}[(\sigma_{\rm{N}}^2{\bm{I}}+P_{\rm{D}}{\bm \Phi})^{-1}{\bm{\widetilde F}}{\bm{\widetilde F}}^{\rm{H}}]}}(\sigma_{\rm{N}}^2{\bm{I}}+P_{\rm{D}}{\bm \Phi})^{-1/2}{\bm{\widetilde F}}_{\rm{opt}}.
\end{align}

Based on \textbf{Conclusion 1}, \textbf{Conclusion 2} and \textbf{Theorem 1}, we define the following variables and parameters
\begin{align}
&\big[\bm{\Lambda}_{\bm{\widetilde F }}^{\rm{T}}\bm{\Lambda}_{\bm{\widetilde F }}\big]_{i,i}=f_i^2, \
\big[\bm{\Lambda}_{\bm{\Psi}}\big]_{i,i}=\psi_i, \nonumber \\
& \big[\bm{\Lambda}_{\bm{X }}\bm{\Lambda}_{\bm{X }}^{\rm{T}}\big]_{i,i}=x_i^2, \
\big[{\bm \Lambda}_{{\bm{R}}_{\rm{N}}}^{-1} \big]_{i,i}=1/\sigma_{i}^2.
\end{align} The joint matrix-monotonic optimization problem
\textbf{P. 17} is equivalent to the following one:
\begin{align}
 \textbf{P. 18:} \  \max_{\{f_i^2\},\{x_i^2\}} \ \ & {\bm{U}}_{\bm{\widetilde F}}\bm{\Lambda}_{\rm{IN}}{\bm{U}}_{\bm{\widetilde F}}^{\rm{H}} \nonumber \\
 \ {\rm{s.t.}} \ & \big[\bm{\Lambda}_{\rm{IN}}\big]_{i,i}= \frac{N_{\rm{R}}^2f_i^2x_i^2 \psi_i/\sigma_i^2}{\sigma_{\rm{N}}^2\psi_i^{-1}+N_{\rm{R}}\sigma_{\rm{N}}^2x_i^2/\sigma_i^2+P_{\rm{D}}} \nonumber \\
& \sum_{i=1}^{N_{\rm{Data}}} f_i^2 \le P_{\rm{D}}, \ \sum_{i}^{N_{\rm{Data}}}x_i^2 \le P_{\rm{T}}T_{\rm{T}},  \nonumber \\
& (T-T_{\rm{T}})P_{\rm{D}}+P_{\rm{T}}T_{\rm{T}}\le E_{\rm{total}}.
\end{align} Based on this, the joint optimization problem \textbf{P. 14} can be further rewritten as
\begin{align}
 \textbf{P. 19:}   \max_{\{f_i^2\},\{x_i^2\},T_{\rm{T}}}  & f_{\rm{unified}}\!\!\left(\!\!\!\left\{\!\! \frac{N_{\rm{R}}^2f_i^2x_i^2 \psi_i/\sigma_i^2}{\sigma_{\rm{N}}^2\psi_i^{-1}\!+\!N_{\rm{R}}\sigma_{\rm{N}}^2x_i^2/\sigma_i^2\!+\!P_{\rm{D}}} \!\! \right\}_{i=1}^{\!\!N_{\rm{Data}}} \!\!\right) \nonumber \\
 & \times \frac{T-T_{\rm{T}}}{T} \nonumber \\
 \ {\rm{s.t.}} \
& \sum_{i=1}^{N_{\rm{Data}}} f_i^2 \le P_{\rm{D}}, \ \sum_{i}^{N_{\rm{Data}}}x_i^2 \le P_{\rm{T}}T_{\rm{T}}, \nonumber \\
&  (T-T_{\rm{T}})P_{\rm{D}}+P_{\rm{T}}T_{\rm{T}}\le E_{\rm{total}}.
\end{align} The specific optimal solutions of \textbf{P. 19} are determined by the particular mathematical formulas of the OF in \textbf{P. 19}. In the following, two specific examples are given to show how to solve \textbf{P. 19}. Their methodology may also be applied to other OFs.

\subsection{Specific Examples for Statistical CSI}

In this subsection, some specific joint optimization problems are investigated one-by-one to show how to jointly optimize the training sequence and the transceiver.

\noindent \textbf{Effective MI Maximization:}

Based on the results derived above, the joint optimization problem \textbf{P. 14}  is finally simplified into the following form for the optimization problem of effective MI maximization
\begin{align}
 \textbf{P. 20:}   \max_{\{f_n^2\},\{x_n^2\},T_{\rm{T}}}  & \!\sum_{i=1}^{N_{\rm{Data}}} \! \log \!\! \left(\!\!1\!+\! \frac{N_{\rm{R}}^2f_i^2x_i^2 \psi_i/\sigma_i^2}{\sigma_{\rm{N}}^2\psi_i^{-1}\!+\!N_{\rm{R}}\sigma_{\rm{N}}^2x_i^2/\sigma_i^2\!+\!P_{\rm{D}}} \!\!\right) \nonumber \\
& \times \frac{T-T_{\rm{T}}}{T} \nonumber \\
  {\rm{s.t.}}
& \sum_{i=1}^{N_{\rm{Data}}} f_i^2 \le P_{\rm{D}}, \ \sum_{i=1}^{N_{\rm{Data}}}x_i^2 \le P_{\rm{T}}T_{\rm{T}}, \nonumber \\
& (T-T_{\rm{T}})P_{\rm{D}}+P_{\rm{T}}T_{\rm{T}}\le E_{\rm{total}}.
\end{align} \textbf{P. 20} can be optimized in an alternating manner with respect to $\{f_n^2\}$, $\{x_n^2\}$, and $T_{\rm{T}}$.
When $T_{\rm{T}}$ is given, for fixed $x_i^2$ values, the optimal solutions of $f_i^2$ are the following water-filling solutions \cite{Boyd04}
\begin{align}
f_i^2=\left( \frac{1}{\mu_f}-\frac{1}{h_i} \right)^{+},
\end{align} where $\mu_f$ is the Lagrange multiplier corresponding to the constraint $\sum_{i=1}^{N_{\rm{Data}}} f_i^2 \le P_{\rm{D}}$.
The parameter $h_i$ is defined as
\begin{align}
h_i=\frac{N_{\rm{R}}^2x_i^2\psi_i/\sigma_i^2} {\sigma_{\rm{N}}^2\psi_i^{-1}+N_{\rm{R}}\sigma_{\rm{N}}^2x_i^2/
\sigma_i^2+P_{\rm{D}}}.\end{align}  Then, for fixed $f_i^2$ values, the optimal solutions of $x_i^2$ are the following water-filling solutions:
\begin{align}
x_i^2\!\!=\!\!\!\left(\!\!\!\frac{-c_i(a_i\!+\!2b_i)\!+\!\!\sqrt{c_i^2(a_i\!\!+\!\!2b_i)^2\!\!-\!\!4(a_i\!\!+\!\!b_i)b_i(c_i^2\!\!-\!\!a_i c_i/\mu_x)}}
{2(a_i\!\!+\!\!b_i)b_i}\!\!\right)^{\!\! +}\!\!,
\end{align} where the parameters $a_i$, $b_i$ and $c_i$ are defined as follows
\begin{align}\label{Definition_a_b_c}
a_i\!=\!N_{\rm{R}}^2f_i^2\psi_i/\sigma_i^2, \
b_i\!=\!N_{\rm{R}}\sigma_{\rm{N}}^2/\sigma_i^2,  \
c_i\!=\! \sigma_{\rm{N}}^2\psi_i^{-1}\!+\!P_{\rm{D}}.
\end{align} Finally, the optimization of $T_{\rm{T}}$ can be carried out using a one-dimensional search\cite{Pastore2016}\cite{Soysal2010}, given its discrete nature. 
{
The solution is illustrated in Algorithm~\ref{alg1}. This procedure is quite similar to the effective MSE minimization.
The outer loop represents the exhaustive search of $T_{\rm{T}}$,
while the inner loop is used for iteratively updating $\bm{X}_{\rm{P}}, \bm{F}$ until the convergence of the OF. The results are averaged over a number of random realizations to find the robust optimal training length and the optimal structure of both the training sequences and of the precoder.}

\begin{algorithm}  [t]
  \caption{ {Joint Effective MI Maximization of Training Sequences and Transceivers Based on Matrix-Monotonic Optimization for Statistical CSI}}
  \label{alg1}
  \begin{algorithmic}
  \STATE {1. Initialize the parameters and begin the exhaustive search from $T_{\rm{T}}$=1. }
  \REPEAT
  \REPEAT
  \STATE {2. Update $f_i^2$ with fixed $x_i^2$ in (56).}
  \STATE {3. Update $x_i^2$ with fixed $f_i^2$ in (58).}
  \UNTIL{the convergence of OF in (55)}
  \STATE {4. $T_{\rm{T}}\leftarrow T_{\rm{T}}+1$.}
  \UNTIL{$T_{\rm{T}}=T-1$}
  \STATE {5. Find the optimal training length $T_{\rm{T}}$, training sequence as in Conclusion 1 and 2, and precoder as in Theorem 1 and (51).}
  \end{algorithmic}
\end{algorithm}

\noindent \textbf{Effective Weighted MSE Minimization}

According to the previous discussions, the joint optimization problem \textbf{P. 14} aims for maximizing a performance metric.
For the effective weighted MSE minimization,
the OF of the joint optimization problem \textbf{P. 14} is formulated as the reciprocal of the effective weighted MSE, i.e.,
\begin{align}
&\frac{T-T_{\rm{T}}}{T}f_{\rm{unified}}\left(\frac{{\bm{ F}}^{\rm{H}}\mathbb{E}\{\bm{\widehat H}^{\rm{H}}{\bm{\widehat H}} \}{\bm{F}}}{\sigma_{\rm{N}}^2+{\rm{Tr}}({\bm \Phi}{\bm{F}}{\bm{F}}^{\rm{H}})}\right)\nonumber \\
=&\frac{T-T_{\rm{T}}}{T}\frac{1}{{\rm{Tr}}\left[\bm{W}\left(\bm{I}+\frac{{\bm{F}}^{\rm{H}}\mathbb{E}\{{\bm{\widehat H}}^{\rm{H}}{\bm{\widehat H}}\}{\bm{F}}}{\sigma_{\rm{N}}^2+{\rm{Tr}}({\bm \Phi}{\bm{F}}{\bm{F}}^{\rm{H}})}\right)^{-1}\right]},
\end{align} based on which the joint optimization problem \textbf{P. 14} with respect to the effective weighted MSE performance finally becomes equivalent to
\begin{align}
 \textbf{P. 21:}   \min_{\{f_n^2\},\{x_n^2\},T_{\rm{T}}}  & \frac{T}{T-T_{\rm{T}}}\sum_{i=1}^{N_{\rm{Data}}}\frac{w_i}{1+ \frac{N_{\rm{R}}^2f_i^2x_i^2\psi_i/\sigma_i^2}{\sigma_{\rm{N}}^2\psi_i^{-1}+N_{\rm{R}}\sigma_{\rm{N}}^2x_i^2/\sigma_i^2+P_{\rm{D}}} } \nonumber \\
 \ {\rm{s.t.}} \ \
& \sum_{i=1}^{N_{\rm{Data}}} f_i^2 \le P_{\rm{D}}, \ \sum_{i=1}^{N_{\rm{Data}}}x_i^2 \le P_{\rm{T}}T_{\rm{T}}, \nonumber \\
& (T-T_{\rm{T}})P_{\rm{D}}+P_{\rm{T}}T_{\rm{T}}\le E_{\rm{total}}.
\end{align}  \textbf{P. 21} can also be optimized in an alternating manner with respect to $\{f_n^2\}$, $\{x_n^2\}$, and $T_{\rm{T}}$.
When $T_{\rm{T}}$ is given, for fixed $x_i^2$ values, the optimal solutions of $f_i^2$ are the following water-filling solutions:
\begin{align}
f_i^2=\left( \sqrt{\frac{w_i}{\mu_f h_i}  }- \frac{1}{h_i}\right)^{+},
\end{align} where $\mu_f$ is the Lagrange multiplier corresponding to the constraint $\sum_{i=1}^{N_{\rm{Data}}} f_i^2 \le P_{\rm{D}}$. On the other hand, for fixed $f_i^2$ values, the optimal solutions of $x_i^2$ are the following water-filling solutions:
\begin{align}
x_i^2=\left(\frac{1}{a_i+b_i}\sqrt{\frac{w_ia_ic_i}{\mu_x}}-\frac{c_i}{a_i+b_i}\right)^{+},
\end{align} where $a_i$, $b_i$ and $c_i$ are defined in (\ref{Definition_a_b_c}). Similar to effective MI maximziation, the optimization of $T_{\rm{T}}$ can be also performed using a one-dimensional search, given its discrete nature.

\subsection{Joint Optimization using the Receiver's Estimated CSI at the Transmitter}
\label{Section_Transmitter_Estimated_CSI}

{
In this section, we investigate the scenario, in which both the transmitter and the receiver have the same estimated CSI, assuming that a feedback channel is available between them. 
The training optimization relies on the statistical information available, while the precoder is optimized based on the channel matrix estimated at the destination and the estimation error model. The final performance metric is averaged over all channel realizations.}
In this case, based on the above mathematical formulation along with our TPC optimization, the joint optimization problem can be further rewritten as follows:
\begin{align}
 \textbf{P. 22:}   \max_{{\bm{F}},{\bm{X}},T_{\rm{T}}}   & \! f_{\rm{unified}}\!\left(\!\frac{{\bm{F}}^{\rm{H}}({\bm{\Psi}}\!-\!{\boldsymbol \Phi})^{\frac{1}{2}}{\bm{H}}_{\rm{W}}^{\rm{H}}{\bm{H}}_{\rm{W}}({\bm{\Psi}}\!-\!{\boldsymbol \Phi})^{\frac{1}{2}}{\bm{F}}}{\sigma_{\rm{N}}^2+{\rm{Tr}}({\boldsymbol \Phi}{\bm{F}}{\bm{F}}^{\rm{H}})}\right)   \nonumber \\
&\times \frac{T-T_{\rm{T}}}{T} \nonumber \\
{\rm{s.t.}}  & \bm{\Phi}=\left({\bm{\Psi}}^{-1}+N_{\rm{R}}{\bm{X}}{\bm{R}}_{\bm{N}}^{-1}
{\bm{X}}^{\rm{H}}\right)^{-1}, \nonumber \\
&{\rm{Tr}}(\bm{F}\bm{F}^{\rm{H}}) \le  P_{\rm{D}}, {\rm{Tr}}(\bm{X}\bm{X}^{\rm{H}}) \le P_{\rm{T}}T_{\rm{T}}, \nonumber \\
& (T-T_{\rm{T}})P_{\rm{D}}+P_{\rm{T}}T_{\rm{T}}\le E_{\rm{total}}.
\end{align} This is equivalent to the following by replacing the sum power constraint ${\rm{Tr}}(\bm{F}\bm{F}^{\rm{H}}) \le  P_{\rm{D}}$ by (\ref{power_constraint_3})
\begin{align}
 \textbf{P. 23:}   \max_{{\bm{F}},{\bm{X}},T_{\rm{T}}}  & \! f_{\rm{unified}}\!\left(\!\frac{{\bm{F}}^{\rm{H}}({\bm{\Psi}}\!-\!{\boldsymbol \Phi})^{\frac{1}{2}}{\bm{H}}_{\rm{W}}^{\rm{H}}{\bm{H }}_{\rm{W}}({\bm{\Psi}}\!-\!{\boldsymbol \Phi})^{\frac{1}{2}}{\bm{F}}}{\sigma_{\rm{N}}^2+{\rm{Tr}}({\boldsymbol \Phi}{\bm{F}}{\bm{F}}^{\rm{H}})}\right)  \nonumber \\
 & \times \frac{T-T_{\rm{T}}}{T} \nonumber \\
 \ {\rm{s.t.}}  & \bm{\Phi}=\left({\bm{\Psi}}^{-1}+N_{\rm{R}}{\bm{X}}{\bm{R}}_{\bm{N}}^{-1}
{\bm{X}}^{\rm{H}}\right)^{-1}, \nonumber \\
&\!\frac{{\rm{Tr}}[\!(\sigma_{\rm{N}}^2{\bf{I}}\!+\!P_{\rm{D}}{\boldsymbol \Phi}){\bm{F}}{\bm{F}}^{\rm{H}}\!]}{\sigma_{\rm{N}}^2+{\rm{Tr}}({\boldsymbol \Phi}{\bm{F}}{\bm{F}}^{\rm{H}})} \!\!\le\! P_{\rm{D}},
{\rm{Tr}}(\!\bm{X}\bm{X}^{\rm{H}}\!) \!\le\! P_{\rm{T}}T_{\rm{T}},
\nonumber \\
& (T\!-\!T_{\rm{T}}\!)P_{\rm{D}}\!+\!P_{\rm{T}}T_{\rm{T}}\!\le\! E_{\rm{total}}.
\end{align}

Similar to \textbf{P. 17}, based on the definition of ${\bm{\widetilde F}}$ in (\ref{Definition_F}), \textbf{P. 23} can  further be reformulated as
\begin{align}
 \textbf{P. 24:}   \max_{{\bm{\widetilde F}},{\bm{X}},T_{\rm{T}}}  & \frac{T-T_{\rm{T}}}{T} f_{\rm{unified}}\big({\bm{\widetilde F}}^{\rm{H}}(\sigma_{\rm{N}}^2{\bf{I}}+P_{\rm{D}}{\boldsymbol \Phi})^{-\frac{1}{2}}({\bm{\Psi}}-{\boldsymbol \Phi})^{\frac{1}{2}}\nonumber \\
 &  \times {\bm{H}}_{\rm{W}}^{\rm{H}}{\bm{H }}_{\rm{W}}({\bm{\Psi}}-{\boldsymbol \Phi})^{\frac{1}{2}}(\sigma_{\rm{N}}^2{\bf{I}}+P_{\rm{D}}{\boldsymbol \Phi})^{-\frac{1}{2}}{\bm{\widetilde F}}\big)  \nonumber \\
 \ {\rm{s.t.}}  & \bm{\Phi}=\left({\bm{\Psi}}^{-1}+N_{\rm{R}}{\bm{X}}{\bm{R}}_{\bm{N}}^{-1}
{\bm{X}}^{\rm{H}}\right)^{-1} \nonumber \\
& {\rm{Tr}}({\bm{\widetilde F}}{\bm{\widetilde F}}^{\rm{H}})\le P_{\rm{D}}, \ {\rm{Tr}}(\bm{X}\bm{X}^{\rm{H}}) \le P_{\rm{T}}T_{\rm{T}}, \nonumber \\
& (T-T_{\rm{T}})P_{\rm{D}}+P_{\rm{T}}T_{\rm{T}}\le E_{\rm{total}}.
\end{align} By exploiting that all elements of $\bm{H}_{\rm{W}}$ are i.i.d. random variables,
\textbf{P. 24} is equivalent to
\begin{align}
 \textbf{P. 25:}   \max_{{\bm{\widetilde F}},{\bm{X}},T_{\rm{T}}}  & \frac{T-T_{\rm{T}}}{T} f_{\rm{unified}}\big({\bm{\widetilde F}}^{\rm{H}}\bm{\Sigma}^{1/2}{\bm{H}}_{\rm{W}}^{\rm{H}}{\bm{H }}_{\rm{W}}\bm{\Sigma}^{1/2} {\bm{\widetilde F}}\big)  \nonumber \\
 \ {\rm{s.t.}} \ \ & \bm{\Sigma}\!=\!(\sigma_{\rm{N}}^2{\bf{I}}+P_{\rm{D}}{\boldsymbol \Phi})^{-\frac{1}{2}}({\bm{\Psi}}\!-\!{\boldsymbol \Phi})(\sigma_{\rm{N}}^2{\bf{I}}+P_{\rm{D}}{\boldsymbol \Phi})^{-\frac{1}{2}}\!,\nonumber \\
 &\bm{\Phi}\!=\!\left({\bm{\Psi}}^{-1}+N_{\rm{R}}{\bm{X}}{\bm{R}}_{\bm{N}}^{-1}
{\bm{X}}^{\rm{H}}\right)^{-1} \nonumber \\
& {\rm{Tr}}({\bm{\widetilde F}}{\bm{\widetilde F}}^{\rm{H}})\le P_{\rm{D}}, \ {\rm{Tr}}(\bm{X}\bm{X}^{\rm{H}}) \le P_{\rm{T}}T_{\rm{T}}, \nonumber \\
& (T-T_{\rm{T}})P_{\rm{D}}+P_{\rm{T}}T_{\rm{T}}\le E_{\rm{total}}.
\end{align}

Based on our matrix-monotonic optimization framework, the OF of \textbf{P. 25} equals
\begin{align}\label{objective_function_distribution}
&f_{\rm{unified}}\big( {\bm{\widetilde F}}^{\rm{H}} \bm{\Sigma}^{1/2}{\bm{H}}_{\rm{W}}^{\rm{H}}{\bm{H }}_{\rm{W}}\bm{\Sigma}^{1/2}  {\bm{\widetilde F}} \big) \nonumber \\
=& f_{\rm{unified}}\left( \Big\{f_i^2\lambda_i(\bm{\Lambda}_{\bm{\Sigma}}^{1/2}{\bm{H}}_{\rm{W}}^{\rm{H}}{\bm{H }}_{\rm{W}}\bm{\Lambda}_{\bm{\Sigma}}^{1/2} )\Big\}_{i=1}^{N_{\rm{Data}}}\right),
\end{align} where the elements of the diagonal matrix  $\bm{\Lambda}_{\bm{\Sigma}}$ are defined as
\begin{align}
\big[\bm{\Lambda}_{\bm{\Sigma}}\big]_{i,i}=\frac{N_{\rm{R}}x_i^2\psi_i/\sigma_i^2} {\sigma_{\rm{N}}^2\psi_i^{-1}+N_{\rm{R}}\sigma_{\rm{N}}^2x_i^2/
\sigma_i^2+P_{\rm{D}}}.
\end{align} The equality in (\ref{objective_function_distribution}) exploits the fact that for any unitary matrix $\bm{U}$, $\bm{H}_{\rm{W}}$ and $\bm{H}_{\rm{W}}\bm{U}$ have the same distribution.
Therefore, based on (\ref{objective_function_distribution})  the joint optimization of our linear TPC and training sequence is formulated as
\begin{align}
 \textbf{P. 26:}   \max_{\{f_i^2\},\{x_i^2\},T_{\rm{T}}} & f_{\rm{unified}}\!\!\left(\!\! \Big\{\!f_i^2\lambda_i(\bm{\Lambda}_{\bm{\Sigma}}^{1/2}{\bm{H}}_{\rm{W}}^{\rm{H}}{\bm{H }}_{\rm{W}}\bm{\Lambda}_{\bm{\Sigma}}^{1/2} )\Big\}_{i=1}^{N_{\rm{Data}}}\right)  \nonumber \\
 & \times \frac{T-T_{\rm{T}}}{T} \nonumber \\
 {\rm{s.t.}}  & \big[\bm{\Lambda}_{\bm{\Sigma}}\big]_{i,i}=\frac{N_{\rm{R}}x_i^2\psi_i/\sigma_i^2} {\sigma_{\rm{N}}^2\psi_i^{-1}+N_{\rm{R}}\sigma_{\rm{N}}^2x_i^2/
\sigma_i^2+P_{\rm{D}}}, \nonumber \\
&  \sum_{i=1}^{N_{\rm{Data}}} f_i^2 \le P_{\rm{D}}, \sum_{i=1}^{N_{\rm{Data}}} x_i^2 \le P_{\rm{T}}T_{\rm{T}}, \nonumber \\
&(T-T_{\rm{T}})P_{\rm{D}}+P_{\rm{T}}T_{\rm{T}}\le E_{\rm{total}}.
\end{align} In the case when the transmitter has estimated CSI, for a given transmit power $P_{\rm{D}}$, the optimal solution of $f_i^2$ can be expressed as a function of $P_{\rm{D}}$ and $\left\{\!\lambda_i(\bm{\Lambda}_{\bm{\Sigma}}^{1/2}{\bm{H}}_{\rm{W}}^{\rm{H}}{\bm{H }}_{\rm{W}}\bm{\Lambda}_{\bm{\Sigma}}^{1/2})\!\right\}$, i.e.,
\begin{align}
f_i^2=p_i\left(P_{\rm{D}},\left\{\lambda_i(\bm{\Lambda}_{\bm{\Sigma}}^{1/2}{\bm{H}}_{\rm{W}}^{\rm{H}}{\bm{H }}_{\rm{W}}\bm{\Lambda}_{\bm{\Sigma}}^{1/2})\right\}_{i=1}^{N_{\rm{Data}}} \right).
\end{align} As a result, the optimization problem \textbf{P. 26} may also be shown to be equivalent to
\begin{align}
 \textbf{P. 27:}   \max_{\{x_i^2\},T_{\rm{T}}}  & f_{\rm{unified}}\left( \left\{f_i^2\lambda_i(\bm{\Lambda}_{\bm{\Sigma}}^{1/2}{\bm{H}}_{\rm{W}}^{\rm{H}}{\bm{H }}_{\rm{W}}\bm{\Lambda}_{\bm{\Sigma}}^{1/2} )\right\}_{i=1}^{N_{\rm{Data}}}\right)  \nonumber \\
 &\times \frac{T-T_{\rm{T}}}{T} \nonumber \\
 {\rm{s.t.}}  & \big[\bm{\Lambda}_{\bm{\Sigma}}\big]_{i,i}=\frac{N_{\rm{R}}x_i^2\psi_i/\sigma_i^2} {\sigma_{\rm{N}}^2\psi_i^{-1}+N_{\rm{R}}\sigma_{\rm{N}}^2x_i^2/
\sigma_i^2+P_{\rm{D}}} \nonumber \\
&f_i^2=p_i\left(P_{\rm{D}},\left\{\lambda_i(\bm{\Lambda}_{\bm{\Sigma}}^{1/2}{\bm{H}}_{\rm{W}}^{\rm{H}}{\bm{H }}_{\rm{W}}\bm{\Lambda}_{\bm{\Sigma}}^{1/2})\right\}_{i=1}^{N_{\rm{Data}}} \right) \nonumber \\
& \sum_{i=1}^{N_{\rm{Data}}} x_i^2 \le P_{\rm{T}}T_{\rm{T}}, \nonumber \\
& (T-T_{\rm{T}})P_{\rm{D}}+P_{\rm{T}}T_{\rm{T}}\le E_{\rm{total}}.
\end{align} It is worth highlighting that the statistical expectation operation involved in \textbf{P. 27} makes the optimization problem difficult to solve. To overcome this difficulty, a pair of algorithms are proposed. For the first one,
the power allocation in the training optimization adopts the suboptimal solutions given by \textbf{P. 1} or \textbf{P. 2}.
As a result, in \textbf{P. 27} there is only a single real scalar optimization variable $T_{\rm{T}}$ that can be optimized by using a simple one-dimensional search. The expectation operation in \textbf{P. 27} can be carried out numerically. On the other hand, the second one is based on the approximations that replace the eigenvalues $\lambda_i(\bm{\Lambda}_{\bm{\Sigma}}^{1/2}{\bm{H}}_{\rm{W}}^{\rm{H}}{\bm{H }}_{\rm{W}}\bm{\Lambda}_{\bm{\Sigma}}^{1/2})$ in \textbf{P. 27} by $\lambda_i(\bm{\Lambda}_{\bm{\Sigma}}^{1/2}\mathbb{E}\{{\bm{H}}_{\rm{W}}^{\rm{H}}{\bm{H }}_{\rm{W}}\}\bm{\Lambda}_{\bm{\Sigma}}^{1/2})$.

\subsection{Specific Examples for Estimated CSI}

\noindent \textbf{Effective MI Maximization:}

In this subsection, we investigate the effective MI maximization in detail, where the MI term in the OF of \textbf{P. 27} equals
\begin{align}
&f_{\rm{unified}}\!\left(\!\! \left\{f_i^2\lambda_i\!\left(\bm{\Lambda}_{\bm{\Sigma}}^{1/2}{\bm{H}}_{\rm{W}}^{\rm{H}}{\bm{H }}_{\rm{W}}\bm{\Lambda}_{\bm{\Sigma}}^{1/2}\right)\!\right\}_{i=1}^{N_{\rm{Data}}}\!\right)\nonumber \\
= & \mathbb{E}\!\left\{\!\sum_{i=1}^{N_{\rm{Data}}} \log\!\!\left(\!1\!+\!f_i^2\lambda_i\!\left(\!\bm{\Lambda}_{\bm{\Sigma}}^{1/2}{\bm{H}}_{\rm{W}}^{\rm{H}}{\bm{H }}_{\rm{W}}\bm{\Lambda}_{\bm{\Sigma}}^{1/2}\!\right)\right)\! \right\}\!.
\end{align}It is widely exploited that for the MI maximization the optimal $f_i^2$ obeys \cite{Palomar03,XingTSP201501}
\begin{align}
f_i^2=\left( \frac{1}{\mu_f}-\frac{1}{\lambda_i(\bm{\Lambda}_{\bm{\Sigma}}^{1/2}{\bm{H}}_{\rm{W}}^{\rm{H}}{\bm{H }}_{\rm{W}}\bm{\Lambda}_{\bm{\Sigma}}^{1/2} )} \right)^{+}.
\end{align} The Lagrange multiplier $\mu_f$ is equivalent to
\begin{align}
\frac{1}{\mu_f}=\frac{P_{\rm{D}}\!+\!\sum_{i=1}^{{\widetilde N}_{\rm{Data}}}\!\frac{1}{\lambda_i(\bm{\Lambda}_{\bm{\Sigma}}^{1/2}
{\bm{H}}_{\rm{W}}^{\rm{H}}{\bm{H }}_{\rm{W}}\bm{\Lambda}_{\bm{\Sigma}}^{1/2} )}}{{\widetilde N}_{\rm{Data}}},
\end{align}where ${\widetilde N}_{\rm{Data}}$ is the number of eigen-channels allocated non-zero powers during the data transmission phase.
Therefore, substituting the value of $\mu_f$ into the OF, the following result holds
\begin{align}\label{Objective_MI_b}
& f_{\rm{unified}}\left( \left\{f_i^2\lambda_i(\bm{\Lambda}_{\bm{\Sigma}}^{1/2}{\bm{H}}_{\rm{W}}^{\rm{H}}{\bm{H }}_{\rm{W}}\bm{\Lambda}_{\bm{\Sigma}}^{1/2} )\right\}_{i=1}^{N_{\rm{Data}}}\right)  \nonumber \\
=&\! \mathbb{E}\!\!\left\{\!\!\sum_{i=1}^{{\widetilde N}_{\rm{Data}}} \!\!\! \log \!\!\! \left(\!\!\!\frac{P_{\rm{D}}\!\!+\!\!\!\sum_{j=1}^{\!{\widetilde N}_{\rm{Data}}}\!\!\!\frac{1}{\lambda_j(\bm{\Lambda}_{\bm{\Sigma}}^{\frac{1}{2}}
{\bm{H}}_{\rm{W}}^{\rm{H}}{\bm{H }}_{\rm{W}}\bm{\Lambda}_{\bm{\Sigma}}^{\frac{1}{2}} )}}{{\widetilde N}_{\rm{Data}}}
\! \lambda_i\!(\!\bm{\Lambda}_{\bm{\Sigma}}^{\frac{1}{2}}{\bm{H}}_{\rm{W}}^{\rm{H}}{\bm{H }}_{\rm{W}}\bm{\Lambda}_{\bm{\Sigma}}^{\frac{1}{2}} \!)\!\!\!\right) \!\!\! \right\}\!\!\!.
\end{align} Based on (\ref{Objective_MI_b}), the effective MI maximization problem is written in the following form
\begin{align}
 \textbf{P. 28:}    \max_{\{x_i^2\},T_{\rm{T}}}   & \mathbb{E}\!\!\left\{\!\sum_{i=1}^{{\widetilde N}_{\rm{Data}}} \! \log \!\left(\frac{P_{\rm{D}}\!+\!\sum_{j=1}^{{\widetilde N}_{\rm{Data}}}\!\frac{1}{\lambda_j(\bm{\Lambda}_{\bm{\Sigma}}^{1/2}
{\bm{H}}_{\rm{W}}^{\rm{H}}{\bm{H }}_{\rm{W}}\bm{\Lambda}_{\bm{\Sigma}}^{1/2} )}}{{\widetilde N}_{\rm{Data}}} \right.\right.\nonumber \\
&\left.\left. \times \lambda_i(\bm{\Lambda}_{\bm{\Sigma}}^{1/2}{\bm{H}}_{\rm{W}}^{\rm{H}}{\bm{H }}_{\rm{W}}\bm{\Lambda}_{\bm{\Sigma}}^{1/2} )\right) \! \right\} \frac{T-T_{\rm{T}}}{T} \nonumber \\
{\rm{s.t.}}  & \sum_{i=1}^{{N}_{\rm{Data}}} \! x_i^2 \! \le \! P_{\rm{T}}T_{\rm{T}},\nonumber \\
& (T\!-\!T_{\rm{T}})P_{\rm{D}}\!+\!P_{\rm{T}}T_{\rm{T}}\!\le\! E_{\rm{total}}, \nonumber \\
& \big[\bm{\Lambda}_{\bm{\Sigma}}\big]_{i,i}=\frac{N_{\rm{R}}x_i^2\psi_i/\sigma_i^2} {\sigma_{\rm{N}}^2\psi_i^{-1}+N_{\rm{R}}\sigma_{\rm{N}}^2x_i^2/
\sigma_i^2+P_{\rm{D}}}.
\end{align} Because of the statistical expectation operation in \textbf{P. 28}, it is challenging to derive the optimal solutions of \textbf{P. 28} in closed forms. In the following two kinds of algorithms are proposed.

In order to avoid complex optimizations involving statistical expectations, some suboptimal solutions are used for $\{x_i^2\}$.
Specifically, since channel estimation is usually performed at high SNRs, for MI maximization, the resources are assumed to be allocated uniformly among $\{x_i^2\}$, i.e., $\{x_i^2=P_{\rm{T}}T_{\rm{T}}/{N_{\rm{Data}}}\}$.
Therefore, \textbf{P. 28} is simplified to the following problem:
\begin{align}
 \textbf{P. 29:}    \max_{T_{\rm{T}}}   & \mathbb{E}\!\!\left\{\!\sum_{i=1}^{{\widetilde N}_{\rm{Data}}}  \log\!\!\left(\!\frac{P_{\rm{D}}\!+\!\sum_{j=1}^{{\widetilde N}_{\rm{Data}}}\!\frac{1}{\lambda_j(\bm{\Lambda}_{\bm{\Sigma}}^{1/2}
{\bm{H}}_{\rm{W}}^{\rm{H}}{\bm{H }}_{\rm{W}}\bm{\Lambda}_{\bm{\Sigma}}^{1/2} )}}{{\widetilde N}_{\rm{Data}}} \right.\right.\nonumber \\ & \left.\left.
\times \lambda_i(\bm{\Lambda}_{\bm{\Sigma}}^{1/2}{\bm{H}}_{\rm{W}}^{\rm{H}}{\bm{H }}_{\rm{W}}\bm{\Lambda}_{\bm{\Sigma}}^{1/2} )\right) \! \right\}
\frac{T-T_{\rm{T}}}{T} \nonumber \\
{\rm{s.t.}}  & x_i^2 \!=\! P_{\rm{T}}T_{\rm{T}}/{N_{\rm{Data}}}, \nonumber \\
&(T\!-\!T_{\rm{T}})P_{\rm{D}}\!+\!P_{\rm{T}}T_{\rm{T}} \!\le\! E_{\rm{total}}, \nonumber \\
& \big[\bm{\Lambda}_{\bm{\Sigma}}\big]_{i,i}=\frac{N_{\rm{R}}x_i^2\psi_i/\sigma_i^2} {\sigma_{\rm{N}}^2\psi_i^{-1}+N_{\rm{R}}\sigma_{\rm{N}}^2x_i^2/
\sigma_i^2+P_{\rm{D}}}.
\end{align}

The second rationale is to replace
the eigenvalues $\lambda_i(\bm{\Lambda}_{\bm{\Sigma}}^{1/2}{\bm{H}}_{\rm{W}}^{\rm{H}}{\bm{H }}_{\rm{W}}\bm{\Lambda}_{\bm{\Sigma}}^{1/2} )$ by their expectations, i.e., by $\lambda_i(\bm{\Lambda}_{\bm{\Sigma}}^{1/2}{\mathbb{E}}\{{\bm{H}}_{\rm{W}}^{\rm{H}}{\bm{H }}_{\rm{W}}\}\bm{\Lambda}_{\bm{\Sigma}}^{1/2} )$.
As a result, \textbf{P. 28} can be approximated as follows:
\begin{align}
 \textbf{P. 30:}    \max_{\{x_i^2\},T_{\rm{T}}}   & \sum_{i=1}^{{\widetilde N}_{\rm{Data}}}  \log\!\!\left(\!\frac{P_{\rm{D}}\!+\!\sum_{j=1}^{{\widetilde N}_{\rm{Data}}}\!\frac{1}{\lambda_j(\bm{\Lambda}_{\bm{\Sigma}}^{1/2}\mathbb{E}\{
{\bm{H}}_{\rm{W}}^{\rm{H}}{\bm{H }}_{\rm{W}}\}\bm{\Lambda}_{\bm{\Sigma}}^{1/2} )}}{{\widetilde N}_{\rm{Data}}} \right.\nonumber \\ & \left.
\times \lambda_i(\bm{\Lambda}_{\bm{\Sigma}}^{1/2}\mathbb{E}\{{\bm{H}}_{\rm{W}}^{\rm{H}}{\bm{H }}_{\rm{W}}\}\bm{\Lambda}_{\bm{\Sigma}}^{1/2} )\right)
\frac{T-T_{\rm{T}}}{T} \nonumber \\
{\rm{s.t.}}  & \sum_{i=1}^{{N}_{\rm{Data}}} \! x_i^2 \! \le \! P_{\rm{T}}T_{\rm{T}}, \nonumber \\
&(T\!-\!T_{\rm{T}})P_{\rm{D}}\!+\!P_{\rm{T}}T_{\rm{T}}\!\le\! E_{\rm{total}}, \nonumber \\
& \big[\bm{\Lambda}_{\bm{\Sigma}}\big]_{i,i}=\frac{N_{\rm{R}}x_i^2\psi_i/\sigma_i^2} {\sigma_{\rm{N}}^2\psi_i^{-1}+N_{\rm{R}}\sigma_{\rm{N}}^2x_i^2/
\sigma_i^2+P_{\rm{D}}}.
\end{align} It is worth noting that this approximation is accurate at high SNRs as shown in Section~\ref{Section_Simulation}, but the proof is omitted due to space limitation. The optimization problem \textbf{P. 30} can be efficiently solved via alternating optimization, where the solution of $\{x_i^2\}$ may be found via the MATLAB function  ``\emph{fmincon}''.

\noindent \textbf{Effective Weighted MSE Minimization:}

In this subsection, the effective weighted MSE minimization is investigated.
For the effective weighted MSE minimization, the MSE term of the OF of \textbf{P. 27} may be expressed as
\begin{align}
&f_{\rm{unified}}\!\left( \!\! \left\{f_i^2\lambda_i(\bm{\Lambda}_{\bm{\Sigma}}^{1/2}{\bm{H}}_{\rm{W}}^{\rm{H}}{\bm{H }}_{\rm{W}}\bm{\Lambda}_{\bm{\Sigma}}^{1/2} ) \!\!\right\}_{i=1}^{{ N}_{\rm{Data}}}\right) \nonumber \\
\!\!=&\! \! \left[\mathbb{E}\left\{\frac{\left(\sum_{i=1}^{{\widetilde N}_{\rm{Data}}}\frac{\sqrt{w_i}}{\sqrt{\lambda_i(\bm{\Lambda}_{\bm{\Sigma}}^{1/2}{\bm{H}}_{\rm{W}}^{\rm{H}}{\bm{H }}_{\rm{W}}\bm{\Lambda}_{\bm{\Sigma}}^{1/2} )}}\right)^2
}{P_{\rm{D}}+\sum_{i=1}^{{\widetilde N}_{\rm{Data}}}\frac{1}{\lambda_i(\bm{\Lambda}_{\bm{\Sigma}}^{1/2}{\bm{H}}_{\rm{W}}^{\rm{H}}{\bm{H }}_{\rm{W}}\bm{\Lambda}_{\bm{\Sigma}}^{1/2} )}}\right\}\!+\!\!\!\!\!\!\sum_{i>{\widetilde N}_{\rm{Data}}}\!\!\!\!\!\omega_i\!\right]^{-1}\!\!\!,
\end{align} based on which \textbf{P. 27} is further reformulated as follows:
\begin{align}
 \textbf{P. 31:}   \!\! \max_{\{x_i^2\},T_{\rm{T}}} \!\!  & \left[ \!\mathbb{E}\!\left\{\!\!\frac{\!\left(\!\sum_{i=1}^{{\widetilde N}_{\rm{Data}}}\!\!\frac{\sqrt{w_i}}{\sqrt{\lambda_i(\bm{\Lambda}_{\bm{\Sigma}}^{1/2}{\bm{H}}_{\rm{W}}^{\rm{H}}{\bm{H }}_{\rm{W}}\bm{\Lambda}_{\bm{\Sigma}}^{1/2} )}}\!\!\right)^{\!\!\!2}
}{P_{\rm{D}}\!\!+\!\!\sum_{i=1}^{{\widetilde N}_{\rm{Data}}}\!\!\frac{1}{\lambda_i(\bm{\Lambda}_{\bm{\Sigma}}^{1/2}{\bm{H}}_{\rm{W}}^{\rm{H}}{\bm{H }}_{\rm{W}}\bm{\Lambda}_{\bm{\Sigma}}^{1/2} )}}\!\!\right\} \!\!+\!\!\!\!\!\!\sum_{i>{\widetilde N}_{\rm{Data}}}\!\!\!\!\!\omega_i\!\right]^{\!\!-\!1} \nonumber \\
& \times \frac{T-T_{\rm{T}}}{T} \nonumber \\
{\rm{s.t.}}  & \sum_{i=1}^{{N}_{\rm{Data}}} \! x_i^2 \! \le \! P_{\rm{T}}T_{\rm{T}}, (T\!-\!T_{\rm{T}})P_{\rm{D}}\!+\!P_{\rm{T}}T_{\rm{T}}\!\le\! E_{\rm{total}}, \nonumber \\
& \big[\bm{\Lambda}_{\bm{\Sigma}}\big]_{i,i}=\frac{N_{\rm{R}}x_i^2\psi_i/\sigma_i^2} {\sigma_{\rm{N}}^2\psi_i^{-1}+N_{\rm{R}}\sigma_{\rm{N}}^2x_i^2/
\sigma_i^2+P_{\rm{D}}}.
\end{align} It is worth noting that solving the optimization problem \textbf{P. 31} is still very challenging due to the statistical expectation operation. In the following, a pair of approximations are used.

Firstly, similar to \textbf{P. 29}, in order to avoid complex mathematical operations, suboptimal power allocation schemes may be applied to $\{x_i^2\}$. At high SNR, the resources are supposed to be allocated proportionally among $\{x_i^2\}$ for sum MSE minimization. Since we are considering the average performance, the suboptimal uniform power allocation scheme is adopted for $\{x_i^2\}$ in the following for simplicity, i.e., $\{x_i^2=P_{\rm{T}}T_{\rm{T}}/{N_{\rm{Data}}}\}$.
Thus, \textbf{P. 31} is simplified to
\begin{align}
 \textbf{P. 32:}   \min_{T_{\rm{T}}}  & \frac{T}{T-T_{\rm{T}}}\mathbb{E}\left\{\frac{\left(\sum_{i=1}^{\widetilde N_{\rm{Data}}}\frac{\sqrt{w_i}}{\sqrt{\lambda_i(\bm{\Lambda}_{\bm{\Sigma}}^{1/2}{\bm{H}}_{\rm{W}}^{\rm{H}}{\bm{H }}_{\rm{W}}\bm{\Lambda}_{\bm{\Sigma}}^{1/2} )}}\right)^2
}{P_{\rm{D}}+\sum_{i=1}^{\widetilde N_{\rm{Data}}}\frac{1}{\lambda_i(\bm{\Lambda}_{\bm{\Sigma}}^{1/2}{\bm{H}}_{\rm{W}}^{\rm{H}}{\bm{H }}_{\rm{W}}\bm{\Lambda}_{\bm{\Sigma}}^{1/2} )}}
\right\}\!\nonumber \\
&+\!\frac{T}{T-T_{\rm{T}}}\!\!\!\!\sum_{i>{\widetilde N}_{\rm{Data}}}\!\!\!{\omega_i}  \nonumber \\
{\rm{s.t.}}  & \big[\bm{\Lambda}_{\bm{\Sigma}}\big]_{i,i}=\frac{N_{\rm{R}}x_i^2\psi_i/\sigma_i^2} {\sigma_{\rm{N}}^2\psi_i^{-1}+N_{\rm{R}}\sigma_{\rm{N}}^2x_i^2/
\sigma_i^2+P_{\rm{D}}}, \nonumber \\
&x_i^2 = \frac{P_{\rm{T}}T_{\rm{T}}}{N_{\rm{Data}}},
 (T\!-\!T_{\rm{T}})P_{\rm{D}}\!+\!P_{\rm{T}}T_{\rm{T}}\!\le\! E_{\rm{total}}.
\end{align} In \textbf{P. 32}, the single optimization variable $T_{\rm{T}}$ can be found by a one-dimensional search.

In the second solution, the eigenvalues $\lambda_i(\bm{\Lambda}_{\bm{\Sigma}}^{1/2}{\bm{H}}_{\rm{W}}^{\rm{H}}{\bm{H }}_{\rm{W}}\bm{\Lambda}_{\bm{\Sigma}}^{1/2} )$  are replaced by their expectations, i.e., by $\lambda_i(\bm{\Lambda}_{\bm{\Sigma}}^{1/2}{\mathbb{E}}\{{\bm{H}}_{\rm{W}}^{\rm{H}}{\bm{H }}_{\rm{W}}\}\bm{\Lambda}_{\bm{\Sigma}}^{1/2} )$. Thus, \textbf{P. 31} is relaxed to
\begin{align}
 \textbf{P. 33:}   \min_{\{x_i^2\},T_{\rm{T}}}  & \frac{T}{T-T_{\rm{T}}}\frac{\left(\sum_{i=1}^{{\widetilde N}_{\rm{Data}}}\frac{\sqrt{w_i}}{\sqrt{\lambda_i(\bm{\Lambda}_{\bm{\Sigma}}^{1/2}
 \mathbb{E}\{{\bm{H}}_{\rm{W}}^{\rm{H}}{\bm{H }}_{\rm{W}}\}\bm{\Lambda}_{\bm{\Sigma}}^{1/2} )}}\right)^2
}{P_{\rm{D}}+\sum_{i=1}^{{\widetilde N}_{\rm{Data}}}\frac{1}{\lambda_i(\bm{\Lambda}_{\bm{\Sigma}}^{1/2}\mathbb{E}\{{\bm{H}}_{\rm{W}}^{\rm{H}}{\bm{H }}_{\rm{W}}\}\bm{\Lambda}_{\bm{\Sigma}}^{1/2} )}}\!\nonumber \\
&+\!\frac{T}{T-T_{\rm{T}}}\!\!\!\!\sum_{i>{\widetilde N}_{\rm{Data}}}\!\!\!{\omega_i} \nonumber \\
{\rm{s.t.}}   & \sum_{i=1}^{N_{\rm{Data}}}\!\! x_i^2 \!\le\! P_{\rm{T}}T_{\rm{T}},
(T\!-\!T_{\rm{T}})P_{\rm{D}}\!+\!P_{\rm{T}}T_{\rm{T}}\!\le\! E_{\rm{total}} \nonumber \\
& \big[\bm{\Lambda}_{\bm{\Sigma}}\big]_{i,i}=\frac{N_{\rm{R}}x_i^2\psi_i/\sigma_i^2} {\sigma_{\rm{N}}^2\psi_i^{-1}+N_{\rm{R}}\sigma_{\rm{N}}^2x_i^2/
\sigma_i^2+P_{\rm{D}}}.
\end{align} The optimization problem \textbf{P. 33} can be solved efficiently by alternating between $\{x_i^2\}$ and $T_{\rm{T}}$, where the solution of $\{x_i^2\}$ may be found by the MATLAB function  ``\emph{fmincon}''.

\subsection{Complexity Analysis}

For the proposed algorithms, the computational complexity mainly arises from the matrix decompositions, water-filling algorithms, numbers of iterations, one-dimensional search over the training interval $T_{\rm{T}}$ and the numerical computations of the statistical expectation.  For an $N\times N$ matrix, the complexity of its matrix decomposition is on the order of $\mathcal{O}(N^3)$. Moreover, for an $N$-dimensional separate optimization problem, the complexity of the water-filling algorithm is as low as $\mathcal{O}(N)$. We will demonstrate in Section~\ref{Section_Simulation} that typically only one or two iterations may be required in Fig.~\ref{Fig_Convergence_MI_Maximization} and Fig.~\ref{Fig_Convergence_MSE_Minimization}.
The one-dimensional search will increase the  computational complexity at an order of $\mathcal{O}(T_{\rm{T}})$. For the following complexity analysis, $L_{\rm{S}}$ denotes the number of independent trials used for the numerical computations of the statistical expectation. Upon using statistical CSI at the transmitter, the total computational complexities of the proposed algorithms are $\mathcal{O}(N^3)+\mathcal{O}(NT_{\rm{T}}+L_{\rm{S}}T_{\rm{T}})$. On the other hand, in the case of estimated CSI at transmitter, the total computational complexities of the proposed algorithms using the eigen-value approximations are also $\mathcal{O}(N^3)+\mathcal{O}(NT_{\rm{T}}+L_{\rm{S}}T_{\rm{T}})$. However, in this case  the total computational complexities of the proposed algorithms using uniform power allocation for training are a little bit different and the total complexity is  $\mathcal{O}(N^3)+\mathcal{O}(L_{\rm{S}}T_{\rm{T}})$.

\section{Simulation Results and Discussions}
\label{Section_Simulation}

\begin{figure}[t]
    \centering
    \includegraphics[width=8.8cm]{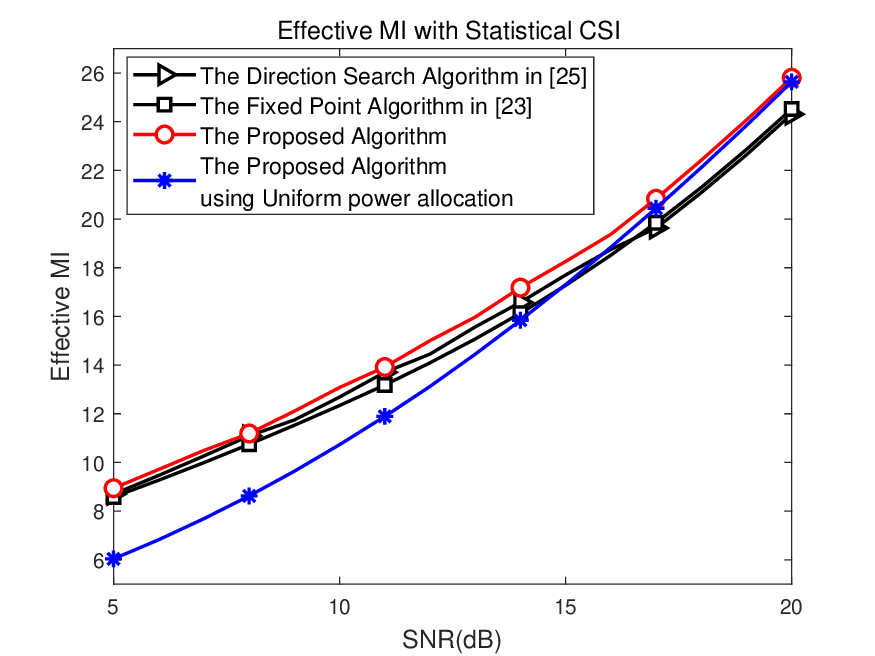}
    \vspace{-3mm}
    \caption{Joint optimization for effective MI maximization using statistical CSI at the transmitter, where the antenna setting is 8$\times$8. }
    \vspace{-5mm}
    \label{Fig_MI_s_Maximization}
\end{figure}

In this section, several numerical results are provided for the proposed joint optimization of our linear TPC and training sequence. Firstly, the  widely used exponential correlation model is used for $\bm{\Psi}$, i.e., $[\bm{\Psi}]_{i,j}=\theta^{|i-j|}$ \cite{XingTSP201501,Ding09}. In the following simulations, $\theta=0.9$ is chosen, and the block length is 256.
For the cases in which the receiver relies on the estimated CSI and the transmitter has only statistical CSI, the joint optimization used for effective MI maximization is investigated first. In Fig.~\ref{Fig_MI_s_Maximization}, the effective MI represents the values of the OF of \textbf{P. 20}, with antenna settings of $N_{\rm{T}}=N_{\rm{R}}=N=8$ and $T=256$ vs the SNR. In order to characterize the performance of the proposed algorithm, the direction search algorithm of \cite{Pastore2016} and  the fixed point algorithm of \cite{Soysal2010} are also used in Fig.~\ref{Fig_MI_s_Maximization}. At low SNRs, the effective MI of the proposed algorithm only has modest advantages over the direction search algorithm of \cite{Pastore2016} and the fixed point algorithm of \cite{Soysal2010}. However, there is a significant performance gap between the proposed algorithm and the algorithm using uniform power allocation, which shows the superiority of the proposed optimization algorithm. With the SNR increasing, the performance gap between the proposed algorithm and the algorithm employing uniform power allocation is reducing, which is due to the fact that the optimal power allocation required for effective MI maximization tends to be uniform at high SNRs. Furthermore, the effective MI of the proposed algorithm is much higher than that of the direction search algorithm of \cite{Pastore2016} and that of the fixed point algorithm of \cite{Soysal2010} at high SNRs.

\begin{figure}[t]
    \centering
    \includegraphics[width=8.8cm]{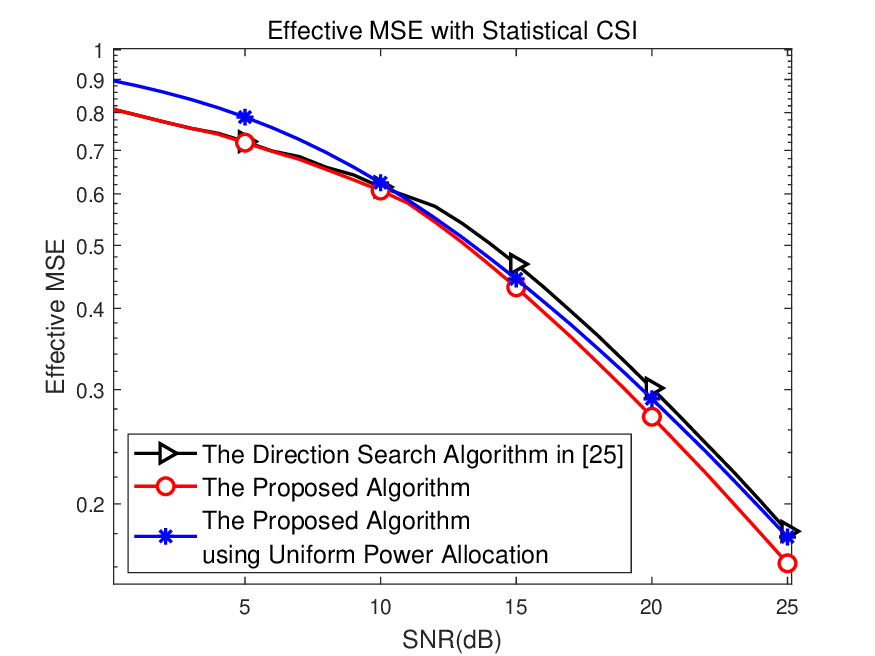}
    \vspace{-3mm}
    \caption{Joint optimization for effective MSE minimization using statistical CSI at the transmitter, where the antenna setting is 8$\times$8. }
    \vspace{-5mm}
    \label{Fig_MSE_s_Minimization}
\end{figure}

In contrast to the MI, the MSE exhibits higher signal recovery accuracy at the receiver. If we directly use MSE as our performance metric, the trivial conclusion emerges that all the resources should be allocated to channel estimation. To overcome this deficiency, the OF of the optimization problem \textbf{P. 21} is the effective MSE instead of the traditional MSE used for transceiver optimization. The effective MSE minimization aims for simultaneously minimizing the data estimation MSE and maximizing the data transmission time interval. In Fig.~\ref{Fig_MSE_s_Minimization}, we compare the effective MSEs of the direction search algorithm \cite{Pastore2016}, of the proposed algorithm, and of the proposed algorithm using uniform power allocation, but only statistical CSI at the transmitter. At high SNRs, the effective MSE of the proposed algorithm is better than that of the direction search algorithm of \cite{Pastore2016}. Additionally, the effective MSE of the proposed algorithm is always lower than that of the proposed algorithm employing uniform power allocation.

\begin{figure}[t]
  \centering
  \includegraphics[width=8.8cm]{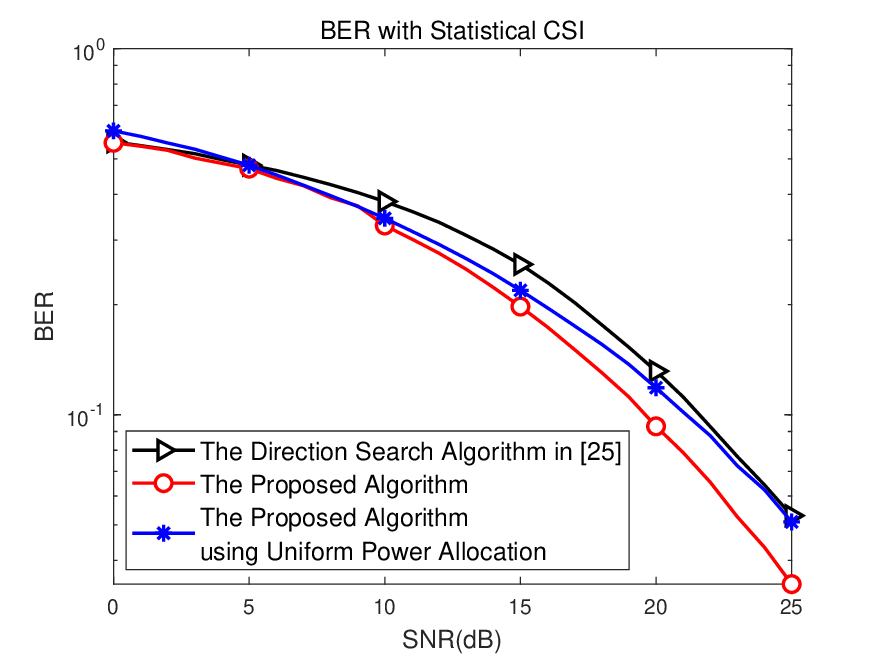}
  \vspace{-3mm}
  \caption{ {The BER performance of joint optimization for effective MSE minimization using statistical CSI at the transmitter, where the antenna setting is 8$\times$8. }}
  \vspace{-5mm}
  \label{ber}
\end{figure}

{
In practical communication systems the bit-error rate (BER) is one of the salient performance criteria. However, its analytical expression is generally quite challenging to derive. Since the mathematically tractable MSE metric is closely related to the BER, Fig.~\ref{ber} illustrates the BER achieved by the MSE solution for the system considered. Observe that at sufficiently high SNRs our solution attains beneficial BER gains over both the direction search algorithm of \cite{Pastore2016} and over that using uniform power allocation.}

\begin{figure}[t]
    \centering
    \includegraphics[width=8.8cm]{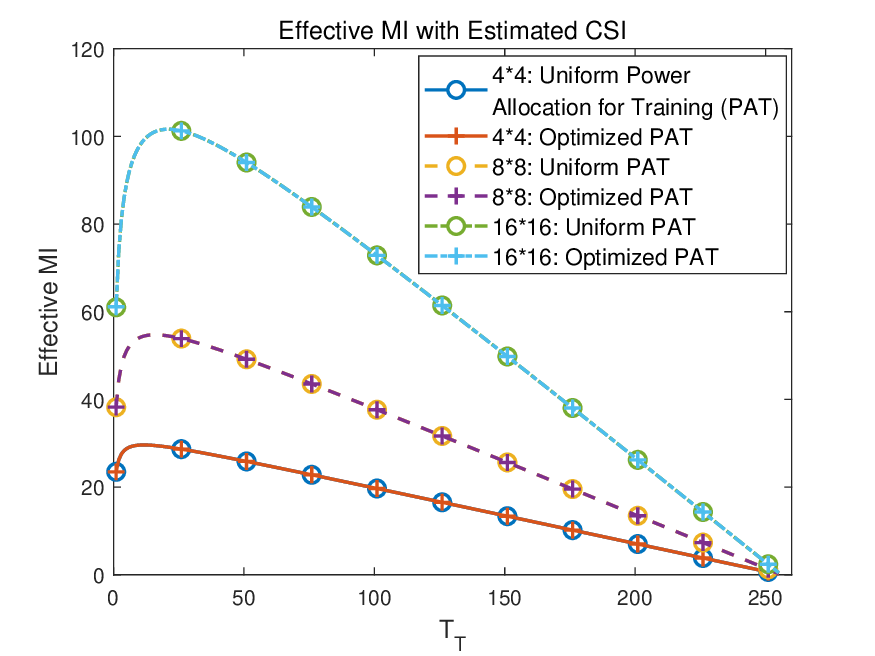}
    \caption{Joint optimization for effective MI maximization using estimated CSI at transmitter with different antenna number settings when SNR is 30 dB. }
    \vspace{-5mm}
    \label{Fig_MI_e_Maximization}
\end{figure}
\begin{figure}[t]
    \centering
    \includegraphics[width=8.8cm]{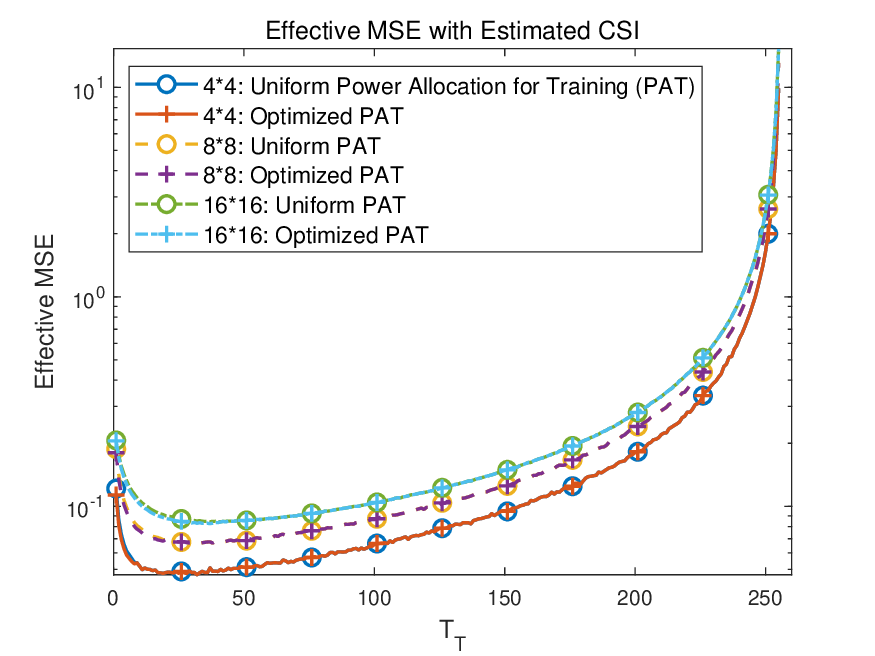}
    \vspace{-3mm}
    \caption{Joint Optimization for effective MSE minimization using estimated CSI at transmitter with different antenna number settings when SNR is 30 dB. }
    \vspace{-5mm}
    \label{Fig_MSE_e_Minimization}
\end{figure}

In the simulations, the other case is also taken into account, in which the transmitter and receiver have the same estimated CSI. In this case, the optimization problems  \textbf{P. 29} and \textbf{P. 30} are used for the effective MI maximization.  In Fig.~\ref{Fig_MI_e_Maximization}, the effective MIs are demonstrated for $T=256$ at SNR$=30$dB and for different antenna settings of $N=4,8,16$. Each point of the curves in Fig.~\ref{Fig_MI_e_Maximization} is an average of $10^4$ independent realizations used for calculating the statistical expectations. It can be concluded from the numerical results that there always exist a best operating point for resource allocation between channel estimation and data transmission. Moreover, it can be seen that for the optimal operating point, the resources allocated to channel estimation are much lower than those allocated for data transmission. Furthermore, Fig.~\ref{Fig_MI_e_Maximization} shows that the suboptimal solution in \textbf{P. 29} using uniform power allocation for training has almost the same performance as the optimized power allocation employed for training in \textbf{P. 30}. Similar conclusions can be drawn for the effective MSE minimization characterized in Fig.~\ref{Fig_MSE_e_Minimization}. For the effective MSE minimization, the optimization problems  \textbf{P. 32} and \textbf{P. 33} are considered. It can also be seen that the suboptimal solution in \textbf{P. 32} associated with uniform power allocation for training has almost the same performance as the optimized power allocation delivered for training in \textbf{P. 33}.

\begin{figure}[t]
    \centering
    \includegraphics[width=8.8cm]{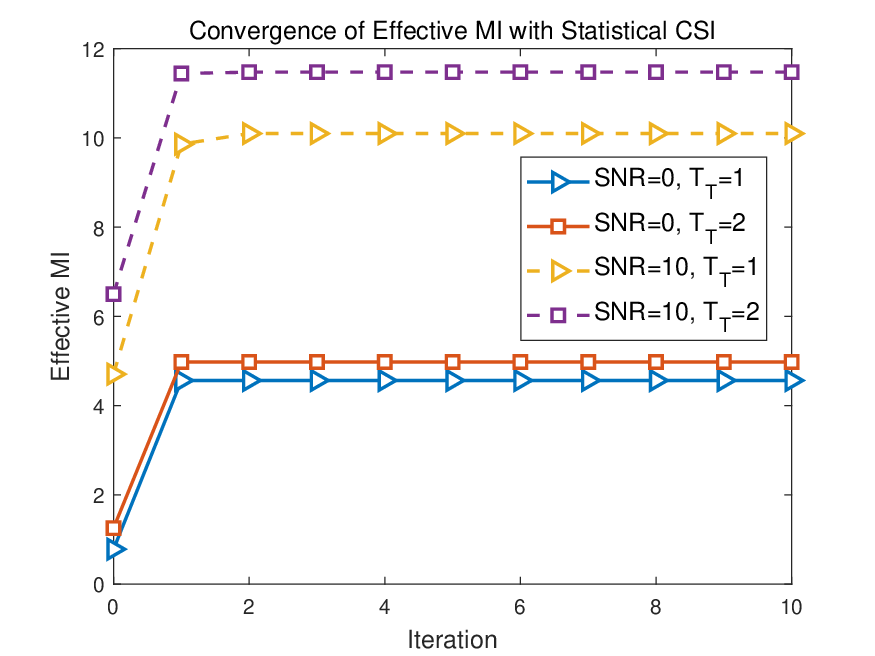}
    \vspace{-3mm}
    \caption{The convergence behavior for effective MI maximization using statistical CSI at transmitter, where the antenna setting is 8$\times$8 under different SNR situations and training lengths. }
    \vspace{-4.5mm}
    \label{Fig_Convergence_MI_Maximization}
\end{figure}

\begin{figure}[t]
    \centering
    \includegraphics[width=8.8cm]{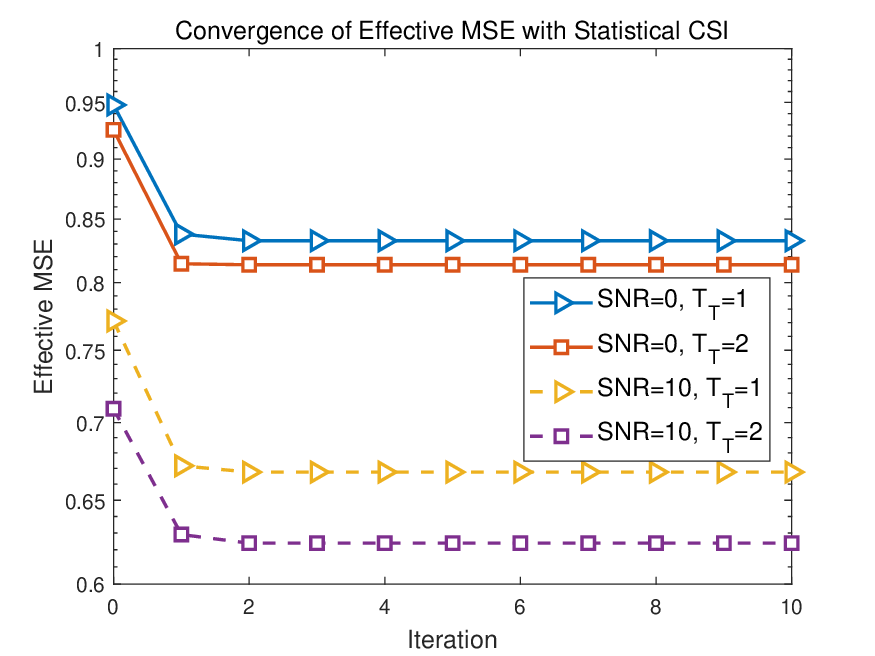}
    \caption{The convergence behavior for effective MSE minimization using statistical CSI at transmitter, where the antenna setting is 8$\times$8 under different SNR situations and training lengths. }
    \vspace{-5mm}
    \label{Fig_Convergence_MSE_Minimization}
\end{figure}

Last but not the least, the convergence behaviors of the proposed algorithms are investigated. It is worth highlighting that for the joint optimization using estimated CSI at the transmitter, the TPC is derived to be a function of the training sequence. In turn, this function is substituted into the original OF. As a result, there is no convergence issue that should be investigated. For the case of statistical CSI at the transmitter, the convergence behaviors of the proposed algorithm are shown in Fig.~\ref{Fig_Convergence_MI_Maximization} and Fig.~\ref{Fig_Convergence_MSE_Minimization}. It is observed that for both effective MI maximization and effective MSE minimization, the proposed algorithms exhibit excellent pretty good convergence properties for all simulation settings.



\section{Conclusions}
\label{Section_Conclusions}

The joint optimization of linear transceivers and training sequences was investigated for various performance metrics, including the effective MI, effective MSE, effective weighted MI, effective weighted MSE, effective additively Schur-convex and  effective additively Schur-concave functions. The new joint matrix-monotonic optimization framework was proposed, which has two matrix variables, i.e., the linear TPC matrix and the training matrix. Based on our new framework of joint matrix-monotonic optimization, the optimal structures of linear transceivers and training sequences were derived for the cases in which the transmitter relies either on statistical CSI or estimated CSI in the face of transmit-side spatial correlation. Based on the optimal structures, the explicit optimization algorithms were proposed for the joint optimization considered. Finally, several numerical simulations were presented for corroborating our theoretical results.

\begin{appendices}
  {
  \section{The Objective Functions for Precoder Optimizations}
  \label{Appendix}
  
In this Appendix, based on the channel model in (\ref{Signal_Model_Data_T}), the signal estimation MSE matrix at the receiver is formulated as:
\begin{align}
\bm{\Phi}_{\rm{MSE}}(\bm{G},\bm{F})\!&=\!\mathbb{E}\left\{\left(\bm{\widehat s}-\bm{s}\right)\left(\bm{\widehat s}-\bm{s}\right)^{\rm{H}}\right\}\nonumber \\
&=\!\mathbb{E}\left\{\left(\bm{G}\bm{y}-\bm{s}\right)\left(\bm{G}\bm{y}-\bm{s}\right)^{\rm{H}}\right\}\nonumber \\
&\succeq \!\bm{\Phi}_{\rm{MSE}}(\bm{G}_{\rm{LMMSE}},\bm{F}),
\end{align} where $\bm{G}_{\rm{LMMSE}}$ is the LMMSE equalizer \cite{Palomar03}. For the MSE matrix $\bm{\Phi}_{\rm{MSE}}(\bm{G}_{\rm{LMMSE}},\bm{F})$,  the $n^{\rm{th}}$ diagonal element reflects the MSE of the $n^{\rm{th}}$ data stream.
As discussed in our previous work \cite{XingTSP201501}, for MIMO precoder/transmitter beamforming optimization, the OFs are  specific matrix functions of the corresponding MSE matrix $\bm{\Phi}_{\rm{MSE}}(\bm{G}_{\rm{LMMSE}},\bm{F})$, which is derived in \cite{XingTSP201501}:
\begin{align}
\bm{\Phi}_{\rm{MSE}}(\bm{G}_{\rm{LMMSE}},\bm{F})=({\bm{I}}\!\!+\!\!{\bm{F}}^{\rm{ H}}{\bm{\widehat H}}^{\rm{H}}{\bm{R}}_{\bm{v}}^{-1}{\bm{\widehat H}}
{\bm{F}})^{-1}.
\end{align}  The matrix ${\bm{R}}_{\bm{v}}$ is the effective noise covariance matrix including both the channel estimation error and the additive noise.
The maximum MI between the transmitter and receiver satisfies $ I_{\max}(\bm{y};\bm{s})\ge -{\rm{log}}\det[\bm{\Phi}_{\rm{MSE}}(\bm{G}_{\rm{LMMSE}},\bm{F})]$ \cite{Pastore2016}. The corresponding  MI maximization OF is
 \begin{align}
\textbf{Obj. 1:} \ {\rm{log}}\det\!({\bm{I}}\!\!+\!\!{\bm{F}}^{\rm{ H}}{\bm{\widehat H}}^{\rm{H}}{\bm{R}}_{\bm{v}}^{-1}{\bm{\widehat H}}
{\bm{F}})\!.
 \end{align}  The sum data MSE is given by the sum of the diagonal elements of $\bm{\Phi}_{\rm{MSE}}(\bm{G}_{\rm{LMMSE}},\bm{F})$ and the corresponding OF of MSE minimization is \cite{XingTSP201501}
 \begin{align}
\textbf{Obj. 2:} \ {\rm{Tr}}[({\bm{I}}\!\!+\!\!{\bm{F}}^{\rm{ H}}{\bm{\widehat H}}^{\rm{H}}{\bm{R}}_{\bm{v}}^{-1}{\bm{\widehat H}}
{\bm{F}})^{-1}\!].
 \end{align} In contrast to the MI, the MSE reflects the recovery accuracy in tangible terms and it is directly related to BER, provided that the error is Gaussian distributed. Moreover, in order to reflect different levels of fairness, a weighted sum performance metric is widely used following the multi-objective optimization framework of \cite{Boyd04}. In contrast to traditional vector valued multi-objective optimizations, the specific performance metric is a matrix-valued function.
 For a positive semidefinite matrix $\bm{\Phi}_{\rm{V}}$,  a general weighting operation may be defined as $\bm{A}_{\rm{W}}^{\rm{H}}\bm{\Phi}_{\rm{V}}\bm{A}_{\rm{W}}+\bm{\Pi}_{\rm{W}}$, with the weighting parameters $\bm{A}_{\rm{W}}$ and $\bm{\Pi}_{\rm{W}}$ being an adjustable complex matrix and positive semidefinite matrix, respectively \cite{WeightedMSEXing}. This kind of weighting operation is much more powerful for matrix-valued multi-objective optimization than traditional element-wise weighting operation. By appropriately varying $\bm{A}_{\rm{W}}$ and $\bm{\Pi}_{\rm{W}}$, the whole Pareto optimal set can be explored.
 Thus, for \textbf{Obj. 1}, taking $\bm{\Phi}_{\rm{V}}={\bm{I}}\!\!+\!\!{\bm{F}}^{\rm{ H}}{\bm{\widehat H}}^{\rm{H}}{\bm{R}}_{\bm{v}}^{-1}{\bm{\widehat H}}
{\bm{F}}$ and defining $\bm{A}=\bm{A}_{\rm{W}}(\bm{A}_{\rm{W}}\bm{A}_{\rm{W}}^{\rm{H}}+\bm{\Pi}_{\rm{W}})^{\frac{1}{2}}$,
 the OF for the weighted MI is written in the following form
  \begin{align}
\textbf{Obj. 3:} \ {\rm{log}}\det\!({\bm{I}}\!\!+\!\!\bm{A}^{\rm{H}}{\bm{F}}^{\rm{ H}}{\bm{\widehat H}}^{\rm{H}}{\bm{R}}_{\bm{v}}^{-1}{\bm{\widehat H}}
{\bm{F}}\bm{A}\!).
 \end{align} Similarly, for \textbf{Obj. 2}, taking $\bm{\Phi}_{\rm{V}}=({\bm{I}}\!+\!{\bm{F}}^{\rm{ H}}{\bm{\widehat H}}^{\rm{H}}{\bm{R}}_{\bm{v}}^{-1}{\bm{\widehat H}}
{\bm{F}}\!)^{-1}$ and defining $\bm{W}=\bm{A}_{\rm{W}}\bm{A}_{\rm{W}}^{\rm{H}}$, the OFs of the weighted MSE is formulated as
 \begin{align}
\textbf{Obj. 4:} \ {\rm{Tr}}[{\bm{W}}({\bm{I}}\!\!+\!\!{\bm{F}}^{\rm{ H}}{\bm{\widehat H}}^{\rm{H}}{\bm{R}}_{\bm{v}}^{-1}{\bm{\widehat H}}
{\bm{F}}\!)^{-1}\!],
 \end{align} where $\bm{W}$ is a positive semidefinite weighting matrix. In the existing literature, a widely used logic for general MIMO transceiver designs is to exploit majorization theory \cite{Palomar03,Xing2021-2}. Based on majorization theory, the OFs are classified into two kinds of functions of the diagonal elements of  $\bm{\Phi}_{\rm{MSE}}(\bm{G}_{\rm{LMMSE}},\bm{F})$ (denoted as ${\bf{d}}[\bm{\Phi}_{\rm{MSE}}(\bm{G}_{\rm{LMMSE}},\bm{F})]$ ), i.e., Schur-convex and Schur-concave functions. The corresponding OFs are expressed as follows
 \begin{align}
\textbf{Obj. 5:} \  f_{{\rm{Concave}}}^{\rm{A\!-\!Schur}}\!\left(\!{\bf{d}}[({\bm{F}}^{\rm{H}}{\bm{\widehat H}}^{\rm{H}}
{\bm{R}}_{{\bm{v}}}^{-1}{\bm{\widehat H}}{\bm{F}}\!\!+\!\!{\bm{I}}\!)^{-1}] \!\right),\nonumber \\
\textbf{Obj. 6:} \  f_{{\rm{Convex}}}^{\rm{A\!-\!Schur}}\!\left(\!{\bf{d}}[({\bm{F}}^{\rm{H}}{\bm{\widehat H}}^{\rm{H}}
{\bm{R}}_{{\bm{v}}}^{-1}{\bm{\widehat H}}{\bm{F}}\!\!+\!\!{\bm{I}}\!)^{-1}] \!\right).
 \end{align} For Schur-convex functions, the performances for different data streams tend to be the same. On the other hand, for Schur-concave functions, the performances of different data streams tend to depend on the eigenvalues of $\bm{\Phi}_{\rm{MSE}}(\bm{G}_{\rm{LMMSE}},\bm{F})$ \cite{Palomar03,Xing2021-2}.
  }
\end{appendices}


\end{document}